\documentclass[12pt]{article}
\usepackage{amsmath}
\usepackage{amssymb}
\usepackage{cite}
\usepackage{float}
\usepackage{latexsym}
\usepackage{color}
\usepackage{graphicx}
\usepackage{caption}
\usepackage{hyperref}
\usepackage{mathtools}
\numberwithin{equation}{section}
\newcommand{\beq}{\begin{equation}}
\newcommand{\be}{\begin{equation}}
\newcommand{\ee}{\end{equation}}
\newcommand{\bea}{\begin{eqnarray}}
\newcommand{\eea}{\end{eqnarray}}

\newcommand{\pa}{\partial}

\newcommand{\nn}{\nonumber}

\newcommand{\mL}{{\mathcal{L}}}
\newcommand{\mc}{\mathcal}

\begin{document}

\pagenumbering{Alph}
\begin{titlepage}
\hbox to \hsize{\hspace*{0 cm}\hbox{\tt }\hss
    \hbox{\small{\tt }}}

\vspace{1 cm}

\centerline{\bf \large Fractional Conformal Descendants and Correlators
 }

\vspace{.6cm}

\centerline{\bf \large  in General 2D $S_N$ Orbifold CFTs at Large $N$}

\vspace{1 cm}
 \centerline{\large
Benjamin A. Burrington$^{\star}$\footnote{benjamin.a.burrington@hofstra.edu}\,,
A.W. Peet$^{\dagger\S}$\footnote{awpeet@physics.utoronto.ca}}

\vspace{0.5cm}

\centerline{\it ${}^\star\!\!$ Department of Physics and Astronomy, Hofstra University, Hempstead, NY 11549, USA}
\centerline{\it ${}^\dagger$Department of Physics, University of Toronto, Toronto, ON M5S 1A7, Canada}
\centerline{\it ${}^\S$Department of Mathematics, University of Toronto, Toronto, ON M5S 2E4, Canada}

\vspace{0.3 cm}

\begin{abstract}
We consider correlation functions in symmetric product ($S_N$) orbifold CFTs at large $N$ with arbitrary seed CFT. Specifically, we consider correlators of descendant operators constructed using both the full Virasoro generators $L_{m}$ and fractional Virasoro generators $\ell_{m/n_i}$.  Using covering space techniques, we show that correlators of descendants may be written entirely in terms of correlators of ancestors, and further that the appropriate set of ancestors are those operators that lift to conformal primaries on the cover.  We argue that the covering space data should cancel out in such calculations.  To back this claim, we provide some example calculations by considering a three-point function of the form (4-cycle)-(2-cycle)-(5-cycle) that lifts to a three-point function of arbitrary primaries on the cover, and descendants thereof.  In these examples we show that while the covering space is used for the calculation, the final descent relations do not depend on covering space data, nor on the details of which seed CFT is used to construct the orbifold, making these results universal.
\end{abstract}

\end{titlepage}

\tableofcontents

\pagenumbering{arabic}


\section{Introduction}

AdS/CFT \cite{Maldacena:1997re,Witten:1998qj} is a remarkable and robust tool for studying questions in quantum gravity, and also provides novel modeling techniques for strongly coupled systems.  The question of which theories of classical gravity may be quantized remains open.  However, one may approach this problem by asking what type of field theories admit gravitational limits.   While this is also not fully answered, some general features are suggested \cite{Heemskerk:2009pn}.  First, the theory must admit a large $N$ limit.  It is clear that for a large $N$ limit to exist the spectrum of operators should be sparse (not scale with $N$) at low conformal dimension.  In addition there should be a large gap controlled by a coupling constant \cite{Heemskerk:2009pn}.

In this work, we concentrate on theories with AdS$_3$ bulk and CFT$_2$ duals.  In these cases, the large $N$ criteria above corresponds to a large central charge $c_{\rm tot}$.  However, simply ``adding degrees of freedom'' is clearly at odds with keeping the spectrum of light operators sparse.  One may consider string theory setups to guide the construction of other models, and the D1/D5 CFT provides a natural starting point.

The D1/D5 system is an extremely well studied setup in string theory and AdS/CFT \cite{Strominger:1996sh,Giveon:1998ns}, providing an instance with an AdS$_3$ bulk, and related black hole backgrounds\cite{David:2002wn}, and a field theory dual.  In this case, one point on the moduli space is argued to be a free orbifold CFT \cite{Dijkgraaf:1998gf,deBoer:1998ip,Seiberg:1999xz,Larsen:1999uk}.  Many features of orbifold models are well understood on general grounds \cite{Dijkgraaf:1989hb}.  In holographic contexts, the orbifold point often plays the role of the free limit of the theory, making it the limit where the CFT is most tractable and where questions regarding coupling constant dependence may be addressed perturbatively.  A general feature in these setups is that the CFT has a large central charge, having a large number of copies of the seed CFT, while the orbifolding process projects out many of the low lying excitations.  Thus, this process generally makes the large $N$ limit plausible. Which orbifold groups are admissible was addressed in \cite{Haehl:2014yla,Belin:2014fna} in the context of permutation orbifolds, where the orbifold group is realized as a subset of permutations acting on $N$ copies of a seed CFT (for further refinements and related ideas, see \cite{Belin:2015hwa,Belin:2016yll,Belin:2017jli}).

Thus, we arrive at a large class of CFTs which admit large $N$ limits: the permutation orbifolds.  Here we mainly focus on the $S_N$ orbifolds, which have been well studied and offer a host of powerful computational methods and interesting results \cite{Dijkgraaf:1996xw,Lunin:2000yv,Lunin:2001pw,Pakman:2009ab, Pakman:2009zz, Pakman:2009mi,Keller:2011xi}.  In addition to this, we must eventually address the deformation of the theory and the introduction of a coupling constant which leads to a large gap \cite{Heemskerk:2009pn}.  For this, computational methods for perturbation theory (and beyond) should be developed for orbifold CFTs.  Developing techniques to compute correlation functions involving the twisted sectors operators is important in this endeavor, and has been undertaken in many examples \cite{Benjamin:2022jin,Guo:2022ifr,Guo:2022sos,AlvesLima:2022elo,Lima:2021wrz,Lima:2020boh,Lima:2020urq,Keller:2019suk,GarciaiTormo:2018vqv,Hampton:2018ygz,Burrington:2017jhh, Carson:2016uwf,Gaberdiel:2015uca,Carson:2014xwa, Carson:2014ena, Carson:2014yxa,Burrington:2014yia,Burrington:2012yq,Avery:2010vk, Avery:2010er, Avery:2010hs}, often in the context of deforming the D1/D5 CFT and related theories where the deformation operator is in the twisted sector \cite{David:1999ec,Lunin:2001pw}.

One may also ask whether there are some universal features which may always appear in permutation orbifold CFTs, regardless of the seed CFT used to construct the permutation orbifold.  Our current aim was prompted in our earlier work \cite{Burrington:2018upk} noting that the universal behavior of the correlators in \cite{Lunin:2000yv} for bare twists at large $N$ suggested the crossing channels be equally universal.  Such a set of crossing channel operators is indeed available, and given by fractional modes of the stress tensor acting on bare twists.  These appear to account for all of the crossing channels \cite{Burrington:2018upk} in this case, at least to the first few orders in crossing channels. This suggests the bare twists and their fractional (and full) Virasoro descendants form a closed subsector of the CFT, at least to leading order at large $N$.  Similar results also exist for the superconformal case \cite{DeBeer:2019oxm}, where members of the superconformal family of the deformation operator were explored.  Other interesting studies addressing fractional modes include \cite{Dei:2019iym,Roumpedakis:2018tdb}, and particularly \cite{Lunin:2001pw} where fractional modes of the superconformal algebra were used to construct the deformation operator of the D1/D5 orbifold CFT starting from a bare twist.  Recently, a similarly universal result was found in \cite{Guo:2022ifr} where they compute the lifting of the conformal dimension of an untwisted superconformal primary in the presence of the deformation operator.

It is interesting to note that many of the above constructions are relatively algebraic, where fractional modes of various currents are used to produce new operators of interest.  Therefore, it seems that some exploration of these fractional excitations is in order.  It is well known that the correlations of descendants using the full Virasoro generators can be written in terms of correlators of primary field ancestors.  It is natural to ask whether, and how, this extends to the fractional excitations as well.  We begin to explore this question here with the most basic of currents in CFTs: the stress tensor.

It is the purpose of this paper to demonstrate that, in the case of symmetric group $S_N$ orbifold CFTs with arbitrary seed CFT at large $N$, the correlation functions of descendants may be written in terms of the correlation functions of an appropriate set of ancestors.  The descendants are constructed using both fractional and full modes of the stress tensor, and the appropriate set of ancestors are those twist sector operators which lift to primaries on the cover (the untwisted sector ancestors also lifts to a set of primaries on the cover, appropriately summed \cite{Burrington:2012yn}, but these are not new considerations).  We use covering space techniques \cite{Lunin:2000yv,Burrington:2018upk} to perform computations, always in the large $N$ limit where the covering space is a sphere.  In the case where we restrict to fractional and full modes of the stress tensor acting on bare twists, this means that these correlation functions may be written in terms of the correlation functions of the ``top'' ancestors, i.e. the bare twists.  The this suggests the fractional and full conformal descendants as a universal closed subsector in orbifold CFTs at large $N$, similar to our earlier work \cite{Burrington:2018upk}.

To begin with, we start with an elementary proof that the correlation functions of the full Virasoro descendants may be written in terms of the correlation function of primary field ancestors.  Although well known, we will generalize this argument to fractional modes later, and so it is worth writing out explicitly here.  We will proceed with an inductive proof.  We consider the correlation function of $Q$ operators
\begin{equation}
\langle \phi_1(z_1) \phi_2(z_2) \phi_3(z_3)\cdots \phi_{Q}(z_{Q})\rangle= f(\{z_q\})
\end{equation}
where the $\phi_q$, $q=1,2,\cdots Q$, are conformal primary fields, i.e.
\begin{equation}
\oint \frac{dz}{2\pi i} (z-z_i)^{m+1} T(z) \phi_i(z_i)=0 \qquad \mbox{for all $m\geq1$}
\end{equation}
(here and throughout, we suppress the antiholomorphic side).   We take that the positions $z_q$ and the conformal dimensions $h_q$ to be given.  When we say that the correlation function of descendants may be written in terms of the ancestors, we mean that they may be written in terms of the given information ($z_q$, $h_q$), and the function $f(\{z_q\})$ (with possible partial derivatives acting on the function).  To proceed, we consider a basis of correlators of descendants as
\begin{equation}
\langle \phi_1^{\{m_{1,i_1}\}}(z_1)\phi_2^{\{m_{2,i_2}\}}(z_2)\phi_3^{\{m_{3,i_3}\}}(z_2) \cdots \phi_Q^{\{m_{Q,i_Q}\}}(z_Q)\rangle \label{Qpoint}
\end{equation}
where we have used the shorthand notation $\{m_{q,i_q}\}$ to be an ordered list of Virasoro generators (adapted to the point in question) acting on the field.  For example
\begin{align}
&\phi_1^{\{-3,-3,-1\}}(z_1)=\left(L_{-3}L_{-3}L_{-1}\phi_1\right)(z_1) \nonumber \\
&=\oint_{z_1}\frac{dz}{2\pi i} \left(z-z_1\right)^{-3+1} T(z)\oint_{z_1}\frac{dz'}{2\pi i} \left(z'-z_1\right)^{-3+1} T(z')\oint_{z_1}\frac{dz''}{2\pi i} \left(z''-z_1\right)^{-1+1} T(z'') \phi_1(z_1).
\end{align}
The indices $m_{q,i_q}$ have a ``double address'', labeling both the operator on which they work, $q$, as well as the position in the product of $L$'s, $i_q$.

It is clear that we may consider all of the $m_{q,i_q}$ to be non-positive because the Virasoro algebra may be used to commute positive indexed $L$'s to the right, and they annihilate the primary fields.  Furthermore, insertions of $L_0$ may be replaced with the eigenvalue of this operator, and so such insertions may be removed knowing the given information $h_q$, and knowing the excitations to the right of the $L_0$ under consideration.   Therefore, we may actually restrict to the case where all $m_{q,i_q}$ are negative (i.e. are excitation operators).  It is further possible to order the operators such that $m_{q,i_q}\leq m_{q,i_q+1}$ using the Virasoro algebra, but this will not come into play for our current proof.

To construct an inductive proof, we consider the integer
\begin{equation}
K_{\rm tot}=-\sum_{q}\sum_{i_q} m_{q,i_q}
\end{equation}
i.e. the total dimension {\it added} to the correlation function by acting with Virasoro generators at all operator insertions.  We will refer to such quantities as the added-dimension, or the dimension-added to a correlator. While the set of correlation functions for a given $K_{\rm tot}$ is combinatorially large, it is still finite.  The inductive assumption is the following: for a given set of $Q$ conformal families (as in (\ref{Qpoint})), we have already proven that all correlators with total dimension-added $K_{\rm tot}\leq K_0$ may be written in terms of the information $z_q, h_q, f(\{z_q\})$.  The start of the induction $K_0=0$ is trivial, where there are no excitations.

We now proceed to show that the inductive assumption implies that all correlators with $K_{\rm tot}\leq K_0+1$ may also be written in terms of the base information $z_q, h_q, f(\{z_q\})$.  It is clear that the only new cases to check are those correlators with dimension-added $K_{\rm tot}=K_0+1$, and we write one such correlator:
\begin{equation}
\langle \phi_1^{\{m_{1,i_1}\}}(z_1)\phi_2^{\{m_{2,i_2}\}}(z_2)\phi_3^{\{m_{3,i_3}\}}(z_2) \cdots \phi_Q^{\{m_{Q,i_Q}\}}(z_Q)\rangle. \label{plus1}
\end{equation}
At least one of the insertions must have excitations acting on it, and without loss of generality, we may assume that $\phi_1$ is one such operator.  Hence $\{m_{1,i_1}\}$ is a non-empty list, and the first entry $m_{1,1}$ is negative.  Focusing in on this term, we write
\begin{equation}
\langle \left[L_{m_{1,1}}\phi_1^{\{m_{1,i_1}\}/m_{1,1}}\right](z_1)\phi_2^{\{m_{2,i_2}\}}(z_2)\phi_3^{\{k_{3,i_3}\}}(z_2) \cdots \phi_Q^{\{m_{Q,i_Q}\}}(z_Q)\rangle.
\end{equation}
where we have used the notation $\{m_{1,i_1}\}/m_{1,1}$ to mean the ordered set of excitations $\{m_{1,i_1}\}$ dropping $m_{1,1}$.

\begin{itemize}
\item {\bf Case 1:} In the case that $m_{1,1}=-1$, then this is just the operator $L_{-1}=\frac{d}{dz_1}$ acting on the field at position $z_1$.  In such a case
\begin{align}
&\langle \left[L_{-1}\phi_1^{\{m_{1,i_1}\}/m_{1,1}}\right](z_1)\phi_2^{\{m_{2,i_2}\}}(z_2) \phi_3^{\{m_{3,i_3}\}}(z_2) \cdots \phi_Q^{\{m_{Q,i_Q}\}}(z_Q)\rangle\nonumber \\
& =\frac{d}{dz_1} \langle \phi_1^{\{m_{1,i}\}/m_{1,1}}(z_1)\phi_2^{\{m_{2,i}\}}(z_2)\phi_3^{\{m_{3,i}\}}(z_2) \cdots \phi_Q^{\{m_{Q,i_Q}\}}(z_Q)\rangle.
\end{align}
We now note that the above is the derivative of a correlation function with total dimension-added $K_{\rm tot}=K_0$.  By the inductive assumption, this correlator may be written in terms of the given information $z_q$, $h_q$, and $f(\{z_q\})$, and so taking the derivative allows us to write the correlator (\ref{plus1}) in terms of $z_q$, $h_q$, and $f(\{z_q\})$ as well, concluding this case.
\item {\bf Case 2:} In the case that $m_{1,1}\leq-2$, we may pull the contour integral defining $L_{m_{1,1}}$.
\begin{align}
&\langle \left[L_{m_{1,1}}\phi_1^{\{m_{1,i_1}\}/m_{1,1}}\right](z_1)\phi_2^{\{m_{2,i_2}\}}(z_2) \phi_3^{\{m_{3,i_3}\}}(z_3) \cdots \phi_Q^{\{k_{Q,i_Q}\}}(z_Q)\rangle\nonumber \\
& =\langle \left[\oint_{z_1} \frac{dz}{2\pi i} (z-z_1)^{m_{1,1}+1} T(z) \phi_1^{\{m_{1,i_1}\}/m_{1,1}}(z_1)\right]\phi_2^{\{m_{2,i_2}\}}(z_2) \phi_3^{\{m_{3,i_3}\}}(z_3) \cdots \phi_Q^{\{k_{Q,i_Q}\}}(z_Q)\rangle.\nonumber \\
& = -\Bigg(\Bigg\langle\phi_1^{\{m_{1,i_1}\}/m_{1,1}}(z_1) \left[\oint_{z_2} \frac{dz}{2\pi i} (z-z_1)^{m_{1,1}+1} T(z) \phi_2^{\{m_{2,i_2}\}}(z_2)\right] \nonumber \\
& \kern16em \times \phi_3^{\{k_{3,i_3}\}}(z_3) \cdots \phi_Q^{\{m_{Q,i_Q}\}}(z_Q)\Bigg\rangle + \cdots\Bigg)
\end{align}
where the contour is pulled to a sum of contours around every other operator: a series of $Q-1$ terms, of which we have only shown the first term above.  One need not worry about pulling through $z=\infty$ because when $z\rightarrow \infty$, the integral becomes $\oint_{\infty} \frac{dz}{2\pi i} (1-z_1/z)^{m_{1,1}+1} z^{m_{1,1}+1} T(z)$.   Expanding the integrand, one can clearly see that all powers of $z$ appearing in the measure are negative, and so represent operators $L_{m_i}$ with negative $m_i$, which annihilate the outgoing vacuum.  Thus the contour pulls through $z=\infty$.

We now note the important point.  The part of the integrand $(z-z_1)^{m_{1,1}+1}$ in the pulled contours $\oint_{z_q} \frac{dz}{2\pi i} (z-z_1)^{m_{1,1}+1} T(z)$ is adapted to the points $z=z_1$ and $z=\infty$, i.e. adapted to certain conventions for the locations of the ``in'' and ``out'' states. When expanded around a different point, $z_q\neq z_1, z_q\neq \infty$, the sequence is a non-singular Laurent series (in fact, a generalized binomial expansion):
\begin{align}
& \oint_{z_q\neq z_1} \frac{dz}{2\pi i} (z-z_1)^{m_{1,1}+1} T(z)\phi_q^{\{m_{q,i_q}\}}(z_q) \nonumber \\
&=\sum_{s=0}^\infty\binom{m_{1,1}+1}{s} (z_q-z_1)^{m_{1,1}+1-s}\oint_{z_q\neq z_1} \frac{dz}{2\pi i}(z-z_q)^s T(z)\phi_q^{\{m_{q,i_q}\}}(z_q) \nn \\
&=\sum_{s=0}^\infty\binom{m_{1,1}+1}{s} (z_q-z_1)^{m_{1,1}+1-s}L_{s-1}\phi_q^{\{m_{q,i_q}\}}(z_q) \label{pulledseries}
\end{align}
Now we can see the resulting correlators have an insertion $\phi_1^{\{m_{1,i_1}\}/m_{1,1}}$ which has lowered the added-dimension by at least $|m_{1,1}|$, i.e. lowered by at least 2.  However, it has been replaced by a sum on terms with $L_{-1}, L_0, L_1, \cdots$ and higher acting on operators at other points in the correlator.  The dimension of these other operators have been raised by at most 1, or possibly left alone, or lowered (possibly substantially).  Thus, the correlators appearing {\it after} the contour pull all have $K_{\rm tot}\leq K_0$, and so fall under the inductive assumption.  Thus, the correlator (\ref{plus1}) with $K_{\rm tot}=K_0+1$ may be written in terms of the information $z_q$, $h_q$, and $f(\{z_q\})$, concluding the proof.
\end{itemize}

It is of further interest to recognize that the series (\ref{pulledseries}) truncates at some finite $s$: the $\phi_q$ represents the lowest conformal dimension operator in some Verma module, and so $L_s$ with $s>-\sum_{i_q} m_{q,i_q}$ annihilates the excited operator.  Thus, the sum over correlators with lower added-dimension is a finite sum. This is of course by necessity because there is a finite number of such correlators.

While the above argument is quite tight, it is useful to summarize the salient (and simple) points that lead to the conclusion.  First, the part of the integrand $(z-z_1)^{m_{1,1}+1}$ defining the left-most excitation acting on the first operator is adapted to two points.  These two points are where this function goes to zero and infinity ($z=z_1,z=\infty$), defining a convention for the location of ``in'' and ``out'' states. When considering excitations, there are only two cases to consider.  The case $m_{1,1}=-1$ is just a derivative.  The case $m_{1,1}\leq -2$ allows for a contour pull, leading to the expansion of $(z-z_1)^{m_{1,1}+1}$ around points for which it is not adapted, e.g. $z_q\neq z_1$, and so $(z-z_1)^{m_{1,1}+1}$ becomes a power series with no poles at the $z_q$'s.  This leads to operators $L_{s}$ with $s\geq -1$ acting on operators at the other points, and so the overall added-dimension in the correlators is lower.  After pulling the contour, the dimension of the $L_{m_{1,1}}$ acting on operator 1, which is now absent, is compensated for with explicit functions of the $z_q$ multiplying lower $K_{\rm tot}$ correlators, and derivatives of such correlators.  A single contour pull leads to the construction of a wanted correlator in terms of a sum of correlators of ancestors (dressed with possible derivatives and functions of locations of the operators).  Iteration of the procedure leads to a desired correlator being written entirely in terms of the correlator of primary operators.

The rest of the paper is organized as follows.  In section 2.1 we establish notation for the full Virasoro generators $L$, the fractional Virasoro generators $\ell$, and the covering space Virasoro generators $\mathcal{L}$.  We establish that the mixed commutator $[L,\ell]$ close on the $\ell$ operators, allowing us to order the $L$ operators to the left.  In section 2.2 we consider how the map to the covering surface establishes relationships between base space operators $L$ and $\ell$, and the covering space operators $\mathcal{L}$.  We show that these sets of operators are reconstructible in terms of each other for fixed added-dimension from the lowest weight state (lowest conformal dimension state), and this establishes that the operators that lift to primaries on the cover are the relevant ancestors to consider when constructing descent relations.  In section 2.3, we provide an inductive proof that the correlators of descendants may be written in terms of ancestors.  While the relationship between the correlators of ancestors and the correlators of descendants naively has dependence on covering space data, we argue that this must cancel, giving answers that only depend on base space data.  We test this expectation in section 3 by computing some examples for three point functions with single cycle twist structure (4)-(2)-(5).  In these examples, we see that descent relations obtained only depend on coarse information: the dimension of operators, twist cycle size, and central charge of the seed CFT, making the results universal.

\section{Generalizing to Fractional Modes}
\label{GenFracModes}
\subsection{Global, Fractional, and Covering space Virasoro Algebras}

Our main concern is the twist operators.  However, we must also consider the untwisted sector because the stress tensor for the orbifold theory is in the untwisted sector.  We write the stress tensor for the orbifold theory as
\begin{equation}
T(z)=\sum_{a=1}^{N} T_a(z)
\end{equation}
where $T_a(z)$ is the stress tensor of the $a^{\rm th}$ copy (we use $a,b,\cdots$ as copy indices).  The modes of $T(z)$ applied to a given operator are as usual
\begin{equation}
(L_{m} \phi_q')(z_q)=\oint_{z_q}\frac{dz}{2\pi i} (z-z_q)^{m+1} T(z) \phi_q'(z_q)
\end{equation}
where $\phi_q'$ is an arbitrary (possibly excited) operator.  This leads to the Virasoro algebra with integer coefficients which satisfies
\begin{equation}
[L_m,L_n]=(m-n)L_{m+n}+\frac{Nc}{12}(m^2-1)m\delta_{m+n,0}.
\end{equation}
Note that we have used $c$ to denote the central charge of the seed CFT, and so for the $S_N$ orbifold CFT the central charge is $Nc$.  We will sometimes refer to the modes $L_{m}$ as the ``full'' Virasoro generators, with corresponding ``full'' algebra.

Now, to consider the twisted sector, we recall that correlation functions are of the form
\begin{equation}
\langle \mathcal{O}_{1,[g_1]} (z_1) \mathcal{O}_{2,[g_2]} (z_2)\cdots \mathcal{O}_{Q,[g_Q]} (z_Q)\rangle
\end{equation}
where $[g_q]$'s above denotes the conjugacy class of the group element $g_q$ which labels the twist of the operator.  These correlators may be written as sums over correlators
\begin{equation}
\langle \mathcal{O}_{1,g_1} (z_1) \mathcal{O}_{2,g_2} (z_2)\cdots \mathcal{O}_{Q,g_Q} (z_Q)\rangle \label{nonGauge1}
\end{equation}
where now the $g_q$ denote a group elements, rather than conjugacy classes, and so now prescribe specific boundary conditions for the copied fields.  While each term is not orbifold invariant, a sum over orbifold images will render it orbifold invariant, and therefore this sum is well defined in the orbifold CFT.   The ordered product of the group elements above must multiply to the identity.

Each of these group elements are products of disjoint cycles $g=({\rm cycle}_1)({\rm cycle}_2)\cdots$.  To each of these disjoint cycles, at each operator, we will construct an associated fractional Virasoro algebra.  Without loss of generality, we may consider the first cycle of the first operator appearing in (\ref{nonGauge1}), and consider this to be the cycle $(1,2,3,\cdots,n)$.  To denote this part of this operator, we use the notation $\sigma_{(1,2,\cdots,n)}'$, where the $'$ reminds us that this twist may be excited in some way.  To excite this operator further, we consider the copies of CFT that are parallel to the cycle, and construct the operator
\begin{equation}
T_{\omega^{-k}}(z)=\sum_{a=1}^n \omega^{k (a-1)}T_a(z)
\end{equation}
where  $\omega=\exp(2\pi i/n)$.  This operator is constructed such that it picks up a phase when circling the operator $\sigma_{(1,2,\cdots,n)}'$ at $z_1$.  We see that $T_a(\exp(2\pi i)(z-z_1)+z_1)=T_{a+1}(z)$ (cyclically), and so
\begin{equation}
T_{\omega^{-k}}\big(\exp(2\pi i)(z-z_1)+z_1\big)=\omega^{-k} T_{\omega^{-k}}(z),
\end{equation}
i.e. the operator is not single valued. However, the combination $\left(z-z_1\right)^{\frac{k}{n}+1}T_{\omega^{-k}}(z)$ is single valued, and so we may construct the modes
\begin{equation}
\ell_{\frac{k}{n}}\sigma_{(1,2,\cdots,n)}' = \oint_{z_1} \frac{dz}{2\pi i} \left(z-z_1\right)^{\frac{k}{n}+1}T_{\omega^{-k}}(z)\sigma_{(1,2,\cdots,n)}'(z_1)
\end{equation}
near this operator.  Just as the factor $(z-z_1)^{m+1}$ in the integral defining $L_m$ is adapted to the point where it acts, so the factor $\left(z-z_1\right)^{\frac{k}{n}+1}$ defining $\ell_{k/n}$ is also adapted to the point where it acts.  However, the $\ell_{k/n}$ have further adaptation through a specification of which cycle the mode acts on, and through this which copies are parallel to that cycle.  Different cycles at the same point in the base space do not interfere, since the different cycles are disjoint, and so involve different copies of the stress tensors.  This appears in covering space computations, where the contours for the $\ell$'s acting on distinct cycles are localized to distinct ramified points in the map $z(t)$, even though these distinct cycles may be part of a single operator in the base space.  To emphasize this, we will mirror our earlier notation for the $L$'s, where they are appended directly to the operator on which they act.  For the $\ell$'s, we append them not only to the operator, but to the specific cycle inside of the operator on which they act.

One may find the commutation relations of the $\ell$ operators acting on a given single cycle twist operator, and one finds
\begin{equation}
\left[\ell_{k/n},\ell_{-k'/n}\right]=\frac{k-k'}{n}\ell_{(k+k')/n}+\delta_{k+k',0} \frac{cn}{12}\left(\left(\frac{k}{n}\right)^2-1\right)\frac{k}{n}.
\end{equation}
which one can show directly using covering space techniques \cite{Burrington:2018upk}.  Now the notation becomes clear.  We use $\ell$, i.e. ``little $\ell$'' to denote these operators because they do not include all copies of the stress tensor, but rather only those copies parallel to the cycle in question.  This becomes manifest in the central charge term appearing as $cn$, i.e. only involving the $n$ copies parallel to the twist.  While this construction by itself is not orbifold invariant, we construct orbifold invariant operators with an appropriate sum over orbifold images.  Finding the correlation functions for the orbifold invariant operators reduces to finding the correlators of these orbifold non-invariant pieces which must then be summed appropriately.

Next, we must be careful to specify the order in which the $L$'s and $\ell$'s are applied, since these do not necessarily commute.  To help us think about this, we consider the covering surface.  One can see that the map $z(t)$ has ramified points for each cycle appearing in \ref{nonGauge1} (although some of these cycles may be part of the same operator on the base, constructing a specific group element $g_q$).  The integrals defining $\ell$'s acting on different cycles within the same operator (in the base space) are lifted to contours around separated points on the cover.  Therefore, it is clear that $\ell$'s attached to distinct cycles do not interfere: the contours are well separated on the cover, so it doesn't make any sense to consider their ordering any more than it makes sense to consider different orderings of $L$'s appended to different operators.  However, the $L$ operators involve {\it all} copies of the stress tensor, including those parallel to the twist.  Thus, the operator $L_{m}=\ell_{m}+L^{\perp}_m$ where the perpendicular part $L^{\perp}$ commutes with any operator parallel to the cycle under considertation.  This makes the commutation relations between $L$'s and $\ell$'s easy to write
\begin{equation}
\left[L_m, \ell_{k/n}\right]=\left(m-\frac{k}{n}\right)\ell_{\frac{k}{n}+m} +\delta_{m+k/n,0} \frac{cn}{12}\left(\left(\frac{k}{n}\right)^2-1\right)\frac{k}{n},
\end{equation}
i.e. the commutation relation between the full Virasoro generators and the fractional Virasoro generators close on the fractional generators.  This relation between the fractional and full-Virasoro generators allows us to pull all of the $L$'s to the left through the $\ell$'s; a simplification we will take advantage of.  Thus, acting with both $L$'s and cycle specific $\ell$'s in any order on one operator, one may use the commutation relations to reduce the operator to a sum of terms of the form (concentrating on the first operator)
\begin{align}
&\Bigg\langle \Bigg(\prod_{i_1=1}^{i_{1,\rm max}} L_{m_{1,i_1}} \times \nonumber \\
&
\qquad \Bigg[
\left(\prod_{i_{1,1}=1}^{i_{1,1,\rm max}}\ell_{k_{1,1,i_{1,1}}/n_{1,1}}\right)\sigma_{(1,2,\cdots,n_{1,1})}' \nonumber \\
&
\qquad \;\; \left(\prod_{i_{1,2}=1}^{i_{1,2,\rm max}}\ell_{k_{1,2,i_{1,2}}/n_{1,2}}\right)\sigma_{(n_{1,1}+1, n_{1,1}+2, \cdots,n_{1,1}+n_{1,2})}' \cdots\Bigg]\Bigg)(z_1) \times \mbox{Other Insertions} \Bigg\rangle \label{bigLpull}
\end{align}
by pulling the $L$'s to the left.

We need to keep track of the index structure quite rigorously here.  If one sees the subscript $m_{3,8}$ one immediately associates that with $L_{m_{3,8}}$ which, through its label, is known to index the $8^{\rm th}$ $L$ operator, acting on the $3^{\rm rd}$ operator insertion in the correlator.  This is why the $m$'s must have ``two addresses'': which operator insertion it works on, and which of the $L$'s in the product the $m$ indexes.  Similarly, the index $k$ must have 3 subscripts.  One denoting the operator insertion, one denoting which cycle within that operator, and one denoting which term in the product of $\ell$'s acting on that cycle.  The $n_{q,j_q}$ must only have two labels: which operator insertion, and which cycle they refer to.  We will continue this notation: $q$ specifies which operator, $j_q$ specifies which cycle.  Now that these are specified, multiple Virasoro operators may be acting on those operators/cycles within operators, and so we introduce the indices $i_{q}$ and $i_{q,j_q}$ to label the $L$ or $\ell$ in this operator one is referring to: the second index on $m_{q,i_{q}}$ and the third index on $k_{q,j_q,i_{q,j_q}}$.  While a bit cumbersome, it makes it easier to keep track of operators and subscripts later on.

This is an important conclusion, and so it is worth summarizing the result.  Excitations using the full-Virasoro and cycle-specific-fractional Virasoro operators, applied in any order, may be written in a basis where the full-Virasoro excitations are written all the way to the left, leaving the cycle-specific-fractional Virasoro excitations to the right.  Since the cycle-specific-fractional Virasoro excitations commute when the cycles are distinct, we may append these generators directly to the cycle twist operator for which they are relevant, as in (\ref{bigLpull}).

There is one final version of the Virasoro algebra that we will need.  The CFT on a connected piece of the covering surface has an associated stress tensor $T(t)$, and the corresponding modes ${\mathcal L}_m$. After the lift, there may be operator insertions at various points $\phi_{\{q,j_q\}\uparrow} (t_{q,j_q})$ on which these modes ${\mathcal L}_m$ act, and these are defined in the usual way
\begin{equation}
{\mathcal L}_m\phi_{\{q,j_q\}\uparrow} (t_{q,j_q})=\oint_{t_{q,j_q}} \frac{dt}{2\pi i} (t-t_{q,j_q})^{m+1} T(t) \phi_{\{q,j_q\}\uparrow} (t_{q,j_q})
\end{equation}
and satisfy the Virasoro algebra on the covering surface
\begin{equation}
[{\mathcal L}_m,{\mathcal L}_n]=(m-n){\mathcal L}_{m+n} + \frac{c}{12}\delta_{m+n,0}(m^2-1)m.
\end{equation}
Note that above the term $c \times 1$ appears in the central charge term because there is only one copy of the seed CFT on the cover.  The modes on the cover ${\mathcal L}$ are of course related to the modes on the base space $L$ and $\ell$ though the lift function $z(t)$.

\subsection{Dimension Reconstruction of Covering Space Virasoro}

Next, we will consider the reconstruction of the covering space Virasoro generators in terms of the covering space generators for the only case we need: operator insertions at ramified points in the map $z(t)$.  For this, we will assume that a certain map $z(t)$ has been specified.  By assuming $z(t)$ is given, the ramified points in the map $t_{q,j_q}$ are automatically specified, although we will often refer to the $t_{q,j_q}$ as being part of the given information.  Each ramified point in the map is associated with a cycle at some given operator insertion, hence the double subscript notation in $t_{q,j_q}$.  Furthermore, if the map $z(t)$ and the points $t_{q,j_q}$ are given, then this specifies the full twist structure, as $z_q=z(t_{q,j_q})$ tells us the location of each cycle twist operator on the base, i.e. which operator the cycle is a part of.  There may be other concerns, for example, which copy indices are common between which cycles, leading to the identity group element, but such global concerns must also be contained in the map $z(t)$ (along with a choice of simply connected patch).  The function $z(t)$ is the information necessary to use the Lunin-Mathur covering space technique to find the correlation function of the bare twists for specified group elements \cite{Lunin:2000yv,Lunin:2001pw,Burrington:2012yn}.

\subsubsection{Lowest Weight Representation Operators on the Cover}

We wish to explore the operator structure appearing at a ramified point in a given map $z(t)$, and without loss of generality, we take this point to be $t_{1,1}$.  In the lift, there will be an operator insertion at the point $t_{1,1}$, which we may break into its conformal families. Because conformal families are orthogonal on the cover, we may consider each conformal family in turn.  We concentrate on the conformal family with lowest conformal dimension operator $\phi_{\uparrow,1,1}(t_{1,1})$.

First, we imagine that some part of an operator $\sigma_{(1,2,\cdots,n_{1,1})}'$ lifts to a finite sum of conformal descendants of $\phi_{\uparrow,1,1}(t_{1,1})$.  We denote this sum of conformal descendants $\phi_{\uparrow,1,1}'(t_{1,1})$.  In such a situation, one can apply the largest annihilation operator $\ell_{k/n_{1,1}}$ to $\sigma_{(1,2,\cdots,n_{1,1})}'$ which does not annihilate it.  Thus,  $\ell_{k/n_{1,1}}\sigma_{(1,2,\cdots,n_{1,1})}'$ is a non-zero operator, and has a lower dimension than $\sigma_{(1,2,\cdots,n_{1,1})}'$.  The lift of this computation is
\begin{equation}
\ell_{k/n_{1,1}}\sigma_{(1,2,\cdots,n_{1,1})}'\rightarrow \oint_{t_{1,1}} \frac{dt}{2\pi i}\frac{(z(t)-z_1)^{k/n_{1,1}+1}}{\pa z(t)}\left(T(t)- \frac{c}{12}\left\{ z(t),t\right\}\right)\phi_{\uparrow,1,1}'(t_{1,1}).
\end{equation}
(here and throughout, we use $\rightarrow$ to mean the computation lifted to the covering surfaces).  The map is adapted at that point, and so has an expansion
\begin{equation}
z(t)=z_1+a_{1,1}(t-t_{1,1})^{n_{1,1}}+\cdots.
\end{equation}
Thus, the lifted computation becomes
\begin{align}
&\oint_{t_{1,1}} \frac{dt}{2\pi i} \frac{a_{1,1}^{\frac{k}{n_{1,1}}}}{n_{1,1}}(t-t_{1,1})^{k+1}\left(1+{\mc O}(t-t_{1,1})\right)
\left(T(t)\right)\phi_{\uparrow,1,1}'(t_{1,1}).
\end{align}
Note, the Schwarzian term has at most a pole of order 2, which appears at the ramified points in the map.  Thus, for $k>0$, this term cannot contribute, and so we have dropped it in the above expansion. The remaining terms on the right are a sum of $\mL_p$'s with $p\geq k$, and so they lower the conformal dimension of all of the terms appearing in $\phi_{\uparrow,1,1}'(t_{1,1})$ by at least $k$.  Thus, the lift of the new operator $\ell_{k/n_{1,1}}\sigma_{(1,2,\cdots,n_{1,1})}'$ contains terms in its lift with lower covering space conformal dimension as well.

One can iterate this process until one reaches a situation where all fractional Virasoro operators with $\ell_{k/n_{1,1}}$ with $k\geq 1$ annihilate the operator $\sigma_{(1,2,\cdots,n_{1,1})}'$.  In such a situation, the lift of these computations states that
\begin{align}
&\oint_{t_{1,1}} \frac{dt}{2\pi i} \frac{a_{1,1}^{\frac{k}{n_{1,1}}}}{n_{1,1}}(t-t_{1,1})^{k+1}\left(1+{\mc O}(t-t_{1,1})\right)
\left(T(t)\right)\phi_{\uparrow,1,1}'(t_{1,1})=0 \label{topOfRep}
\end{align}
for all $k>0$. Remember, the operator $\phi_{\uparrow,1,1}'(t_{1,1})$ is a finite sum of covering space descendants of $\phi_{\uparrow,1,1}(t_{1,1})$.  Therefore, there is some value of $n_{\rm max}$ such that $\mL_n\phi_{\uparrow,1,1}'(t_{1,1})=0$ for all $n>n_{\rm max}$.  In such a situation, we may look at (\ref{topOfRep}), and realize that $k=n_{\rm max}$ does not need to be expanded at all: the higher terms lead to $\mL_n$ acting on $\phi_{\uparrow,1,1}'(t_{1,1})$ with $n>n_{\rm max}$.  Thus, assuming $n_{\rm max}\geq 1$, we plug in $k=n_{\rm max}$ and find
\begin{align}
&\oint_{t_{1,1}} \frac{dt}{2\pi i} \frac{a_{1,1}^{\frac{n_{\rm max}}{n_{1,1}}}}{n_{1,1}}(t-t_{1,1})^{n_{\rm max}+1}
T(t)\phi_{\uparrow,1,1}'(t_{1,1})=0
\end{align}
in contradiction to the statement that $\mL_{n_{\rm max}}\phi_{\uparrow,1,1}'(t_{1,1})$ is non 0.  Thus, $n_{\rm max}=0$, and so  $\phi_{\uparrow,1,1}'(t_{1,1})$ is proportional to $\phi_{\uparrow,1,1}(t_{1,1})$, i.e. is a primary operator once lifted to the cover.

We state this more succinctly. Consider a base space operator, lifted to the covering surface.  If any combination of conformal descendants of $\phi_{\uparrow,1,1}(t_{1,1})$ appears in the lift, then there exists another operator in the base where the lift of that operator is proportional to the primary $\phi_{\uparrow,1,1}(t_{1,1})$.  This is because the statement $\ell_{k/n} \sigma'=0$ for all $k\geq 1$ necessarily implies that ${\mc L}_k \phi_{\{q,j_q\}\uparrow}=0$ for all $k\geq 1$.  Initially, the use of $\ell_{k/n}$ with $k\geq 1$ to ``step up'' through the ancestors has made a list of operators that each make sense on the base.  Thus, the final non-zero operator constructed also makes sense as a base space operator.

\subsubsection{Weight Reconstruction of Descendants}

We have now established that any operator under consideration has an ancestor which lifts to a primary on the cover.  Algebraically, these are defined via $L_m\phi=0$ and $\ell_{m/n_i}\phi=0$ for all $m>0$ for all full $L$ operators, and all cycle specific $\ell$ operators.  These conditions are consistent because the commutation relations, and because the weights of such operators are additive.  Now we wish to show that any descendant on the cover may be constructed by using fractional descendants in the base space.

We begin by considering an operator which lifts to a primary on the cover
\begin{equation}
\underline{\sigma}_{(1,2,\cdots,n_{1,1})}\rightarrow \phi_{\uparrow,1,1}(t_{1,1}).
\end{equation}
Note, we reserve the notation $\underline{\sigma}$ for twists which lift to primaries.  We consider a product of excitations on the base space operator
\begin{equation}
\left(\prod_{i_{1,1}=1}^{i_{1,1,\rm max}}\ell_{k_{i_{1,1}}/n_{1,1}}\right)\underline{\sigma}_{(1,2,\cdots,n_{1,1})}
\end{equation}
this lifts to an insertion of
\begin{equation}
\rightarrow \prod_{i_{1,1}=1}^{i_{1,1,\rm max}}\left[\oint_{t_{1,1}} \frac{dt}{2\pi i}\frac{(z(t)-z_1)^{\frac{k_{i_{1,1}}}{n_{1,1}}+1}}{\pa z(t)}\left(T(t)- \frac{c}{12}\left\{ z(t),t\right\}\right)\right]\phi_{\uparrow,1,1}(t_{1,1}) \label{baseProduct}
\end{equation}
where $\phi_{\uparrow,1,1}(t_{1,1})$ is a primary on the cover.  Thus, the above product of base space $\ell$'s may be written in terms of sums of products of $\mL$'s on the cover.  We wish to show that the reverse is also true.

Paying attention to the expansion of this operator, the highest dimension string of $\mL$ operators arises from keeping only the most singular part of the measure for each of the contour integrals.  Thus, the above computation lifts to
\begin{equation}
C^{\{k_{i_{1,1}}\}}\prod_{i_{1,1}=1}^{i_{1,1,\rm max}}\mL_{k_{i_{1,1}}}\phi_{\uparrow,1,1}(t_{1,1})+\cdots
\end{equation}
where $C^{\{k_{i_{1,1}}\}}$ is a map-data dependent coefficient (involving $a_{1,1}$, e.g.), and the $\cdots$ represents lower covering space dimension terms.  Thus, to reconstruct an arbitrary sum over covering space conformal descendants of $\phi_{\uparrow,1,1}(t_{1,1})$, one simply organizes them by dimension.  The highest dimension operators may be replaced by $1/C^{\{k_{i_{1,1}}\}}$ times the expression (\ref{baseProduct}). This takes care of the highest dimension operators, but disturbs lower dimension operators.  However, the next lowest dimension operators may then be replaced similarly, and so on.  Eventually, the entire sum over descendants has been reconstructed, using the covering space dimension to organize the calculation.  This leads to the following inductive proof.

We consider the covering space conformal descendants of $\phi_{\uparrow,1,1}(t_{1,1})$. All such descendants may be written using a basis
\begin{align}
\prod_{i_{1,1}=1}^{i_{1,1,\rm max}}\mL_{k_{i_{1,1}}}\phi_{\uparrow,1,1}(t_{1,1}) \label{coverGenDes}
\end{align}
where all $k_{i_{1,1}}<0$.  We define the covering space added-dimension at $t_{1,1}$ as
\begin{equation}
K^{\uparrow}_{1,1}=-\sum_{i_{1,1}} k_{i_{1,1}}.
\end{equation}
The inductive assumption is that all operators on the cover (\ref{coverGenDes}) with $K^\uparrow_{1,1}\leq K^\uparrow_0$ may be written in terms of the lift of a sum of operators of the form
\begin{equation}
\left(\prod_{i_{1,1}=1}^{i_{1,1,\rm max}}\ell_{k_{i_{1,1}}/n_{1,1}}\right)\underline{\sigma}_{(1,2,\cdots,n_{1,1})}. \label{arbFrac}
\end{equation}
Again the $K^\uparrow_0=0$ case is trivial to start the induction.  We consider an arbitrary monomial term with $K^\uparrow_{1,1}=K^\uparrow_0+1$ on the cover,
\begin{equation}
\prod_{i_{1,1}=1}^{i_{1,1,\rm max}}\mL_{k_{i_{1,1}}}\phi_{\uparrow,1,1}(t_{1,1}). \label{monomialLift}
\end{equation}
We see that
\begin{align}
& \frac{1}{C^{\{k_{i_{1,1}}\}}}\left(\prod_{i_{1,1}=1}^{i_{1,1,\rm max}}\ell_{k_{i_{1,1}}/n_{1,1}}\right)\underline{\sigma}_{(1,2,\cdots,n_{1,1})} \nonumber \\
& \rightarrow \prod_{i_{1,1}=1}^{i_{1,1,\rm max}}\mL_{k_{i_{1,1}}}\phi_{\uparrow,1,1}(t_{1,1})+{\mc C}_{LD}
\end{align}
where ${\mc C}_{LD}$ is a covering space operator constructed from lower dimension terms, i.e. those with $K^\uparrow_{1,1}\leq K^\uparrow_0$.  Under the inductive assumption these may be written as the lift of a sum of operators (\ref{arbFrac}), i.e. there exists a base space operator $\mathcal{B}_{LD}$ written as a sum of terms of the form (\ref{arbFrac}) such that
\begin{equation}
\mathcal{B}_{LD}\rightarrow {\mc C}_{LD}
\end{equation}
and so
\begin{align}
& \frac{1}{C^{\{k_{i_{1,1}}\}}}\left(\prod_{i_{1,1}=1}^{i_{1,1,\rm max}}\ell_{k_{i_{1,1}}/n_{1,1}}\right)\underline{\sigma}_{(1,2,\cdots,n_{1,1})}- \mathcal{B}_{LD} \nonumber \\
& \rightarrow \prod_{i_{1,1}=1}^{i_{1,1,\rm max}}\mL_{k_{i_{1,1}}}\phi_{\uparrow,1,1}(t_{1,1}),
\end{align}
and we have reconstructed (\ref{monomialLift}), completing the proof.  At this step it is important to note that the construction necessarily contains the coefficient $C^{\{k_{i_{1,1}}\}}$, which generically contains map information.  However, it is exactly the map dependence of these coefficients (and those appearing in subleading terms) that combine with covering space $n$-point functions and lead to cancellations, as we will see in the examples.

In fact, we have show that the fractional Virasoro algebra attached to the first cycle of the first operator can be mapped via an invertible map onto the Virasoro algebra on the cover for specific lowest weight modules on the cover.  It is worth mentioning that this correspondence matches the highest dimension operators in each expansion, schematically
\begin{align}
& \left(\prod_{i_{1,1}=1}^{i_{1,1,\rm max}}\ell_{k_{i_{1,1}}/n_{1,1}}\right)\underline{\sigma}_{(1,2,\cdots,n_{1,1})} +\cdots \rightarrow C^{\{k_{i_{1,1}}\}}\prod_{i_{1,1}=1}^{i_{1,1,\rm max}}\mL_{k_{i_{1,1}}}\phi_{\uparrow,1,1}(t_{1,1})\nonumber +\cdots
\end{align}
where the $\cdots$ on each side represent smaller dimension ladder operators on each side.  The maximum dimension operators therefore have related total added-dimension.  Defining $K_{1,1}=-\sum_{i_{1,1}} k_{i_{1,1}}/n_{1,1}$ for the base space added-dimension, we see that $K^{\uparrow}_{1,1}=n_{1,1} K_{1,1}$, for the largest dimension operators in each expansion.

The content of the last two sections also shows that any operator that lifts to primaries on the cover is a primary in the base space.  If we apply $L_{m}$ with $m>0$ on any operator that lifts to primaries on the cover, then the leading order terms in the cover are $\mL_{n_{q,j_q}m}$ at any point of interest (even unramified points, where one can set $n_{q,j_q}=1$ in the above expression).  Thus, the leading order term annihilates the operator on the cover, as do all subsequent terms, and so the operator in the base space is a primary.  It is, however, not true that any primary operator lifts to a primary on the cover.  The class of primary operators that lift to primaries on the cover is a special class of primaries: those primaries that satisfy not only $L_m\phi=0$ for $m>0$, but also satisfy $\ell_{m/n_i}\phi=0$ for all $m>0$ for all cycle specific $\ell$ operators as well.

It may be tempting to think of the above as isomorphisms of the universal enveloping algebras.  However, this is not really the case.  Rather, it is an isomorphism of the vectors (i.e. the module) the operators act on.  To find the map between a given product of $\mL$'s and a sum of product of $\ell$'s, one must specify ``how deep'' one has excited the lowest conformal dimension ancestor, i.e. which vector in the multiplet it is working on.  One may, therefore, think of this only as an isomorphism between the excitation sector of the two algebras acting on a given lowest conformal dimension operator.  The sums involved may only be calculable for fixed ``depth'', where the sums are necessarily finite.  For higher dimension descendants, more terms are involved in the map between the universal enveloping algebras.

\subsection{Proof for fractional modes and descendants}
\label{fracmoddecend}

We now wish to show that correlators of descendants, both full and fractional descendants, may be written in terms of ancestors in some way.  The most general form of the correlators we will consider is
\begin{align}
&\Bigg\langle \Bigg(\prod_{i_1=1}^{i_{1,\rm max}} L_{m_{1,i_1}} \times \nonumber \\
&
\qquad \Bigg[
\left(\prod_{i_{1,1}=1}^{i_{1,1,\rm max}}\ell_{k_{1,1,i_{1,1}}/n_{1,1}}\right)\underline{\sigma}_{(1,2,\cdots,n_{1,1})} \nonumber \\
&
\qquad \;\; \left(\prod_{i_{1,2}=1}^{i_{1,2,\rm max}}\ell_{k_{1,2,i_{1,2}}/n_{1,2}}\right)\underline{\sigma}_{(n_{1,1}+1, n_{1,1}+2, \cdots,n_{1,1}+n_{1,2})} \cdots\Bigg]\Bigg)(z_1) \times \mbox{Other Insertions} \Bigg\rangle \label{bigLpullBase}.
\end{align}
with the ``Other Insertions'' being of the same form, simply at other points $z_q$.
The only difference between (\ref{bigLpullBase}) and (\ref{bigLpull}) is that the cycle twist operators have the prime dropped in (\ref{bigLpullBase}), and replaced with $\underline{\sigma}$ where this implies that these twists are lifted to primaries on the cover (which we have shown is possible in previous steps).  Again, we emphasize that $m_{q,i_q}<0$, and $k_{q,q_j,i_{q,j_q}}<0$.  One may also choose a basis such that all terms appearing above have $m_{q,i_q}\leq m_{q,i_q+1}$, and $k_{q,j_q,i_{q,j_q}}\leq k_{q,j_q,i_{q,j_q}+1}<0$ such that the excitations are ordered with highest dimension operators to the left.

First, we will show that the correlators with full Virasoro generators and fractional Virasoro generators can be written in terms of correlators only involving the fractional generators and derivatives of such terms.  This will simplify matters by allowing us to pay attention only to the connected covering surface when examining a lift.  If full Virasoro excitations remain, then the excitations have parts parallel to twist directions that appear in cycles in the correlator (at any point), and components perpendicular to all cycles involved in the correlator (requiring us to keep track of a host of these as well).

We again seek an inductive proof.  We consider the total dimension-added to the correlator
\begin{equation}
K_{\rm tot}=-\sum_q \left(\sum_{i_q}m_{q,{i_q}}+\sum_{j_{q}}\sum_{i_{q,j_q}} \frac{k_{q,j_q,i_{q,j_q}}}{n_{q,j_q}}\right).
\end{equation}
The sum on $q$ is obvious: this is a sum over the operator insertions on the base space.  The first term inside the $q$ sum is the total dimension-added by full Virasoro generators, while the second term is the total dimension-added by fractional excitations, summed cycle by cycle.  However, the above is not an integer, since the dimensions are fractional.  However, if we take the least common multiple of the $\{n_{q,j_q}\}$ to be
\begin{equation}
n_{\rm lcm}={\rm lcm}(\{n_{q,j_q}\})
\end{equation}
then the number
\begin{equation}
n_{\rm lcm} K_{\rm tot}=\hat{K}_{\rm tot}
\end{equation}
is an integer, and we can construct an inductive proof using this integer.

The inductive assumption is that all correlators of the form (\ref{bigLpullBase}) with the given group structure (specifiying the $n_{q,j_q}$'s) with $\hat{K}_{\rm tot}\leq \hat{K}_0$ may be written as sums of correlators of the form (\ref{bigLpullBase}) with no full Virasoro generators appearing, and derivatives of such terms.  The start of the induction $\hat{K}_0=0$ is trivial, as always, as this implies no excitations at all.

So, now we consider an arbitrary correlator of the form (\ref{bigLpullBase}) with $\hat{K}_{\rm tot} = \hat{K}_0+1$.  There are three cases. {\bf Case 1} is the case where there are no operators with full Virasoro excitations, which is trivial: it is already of the form (\ref{bigLpullBase}) with no full Virasoro generators appearing.  {\bf Case 2} is where one operator has $m_{q,i_q}=-1$ for $i_q=1$ at any point i.e. an $L_{-1}$ is present as the left most term at any site $z_q$.  In such a case, this is just the derivative of a correlator with $\hat{K}_{\rm tot} \leq \hat{K}_0$, and so the correlator can be written in the form (\ref{bigLpullBase}) and derivatives of such terms.  The only remaining case is {\bf Case 3}, where at least one site must contain full Virasoro excitations, and at least one of these sites contains full Virasoro excitations with the leftmost operator $L_{m_{q,i_q}}$ with $m_{i_{q}}\leq -2$.  Without loss of generality, we assume this is the first operator.

In such a situation, we may pull the integral defining $L_{m_{1,1}}$, arriving at a sum of terms of the form
\begin{align}
&\sum_{q=2}^{Q}-\Bigg\langle\left(\mbox{first insertion stripped of $L_{m_{1,1}}$} \right) \cdots \nonumber \\
& \oint_{z_q} \frac{dz}{2\pi i} (z-z_1)^{m_{1,1}+1} T(z) \Bigg(\prod_{i_q=1}^{i_{q,\rm max}} L_{m_{q,i_q}} \times \nonumber \\
&
\qquad \Bigg[
\left(\prod_{i_{q,1}=1}^{i_{q,1,\rm max}}\ell_{k_{q,1,i_{q,1}}/n_{q,1}}\right)\underline{\sigma}_{{{\rm cycle}_{q,1}}} \nonumber \\
&
\qquad \;\; \left(\prod_{i_{q,2}=1}^{i_{q,2,\rm max}}\ell_{k_{q,2,i_{q,2}}/n_{q,2}}\right)\underline{\sigma}_{({\rm cycle}_{q,2})} \cdots\Bigg]\Bigg)(z_q) \times \mbox{Other Insertions} \Bigg\rangle \nonumber
\end{align}
We note that the dimension of the first operator has been lowered by at least 2.  The other terms in the sum have integrals over $T(z)$ with a measure not adapted to the point, and so become non-singular Laurent series, and so are sums of terms with $L_{m}$ with all $m>-1$.  Thus, the dimension of these terms has been increased by at most 1.  Thus, $\hat{K}_{\rm tot}$ has decreased by at least $n_{\rm lcm}$ in each of these correlators, and so the inductive assumption applies to all of these terms.  Remember, one can always push $L_m$ with $m>0$ in to the right using commutators, and the resulting operators in the sum will still be of the form $L$'s to the left, and $\ell$'s to the right because the $[L,\ell]$ commutators close on the $\ell$'s, and $L_m$ with with $m>0$ annihilate operators that lift to primaries.  In spirit, the proof above is identical to the one in the introduction.  The only new piece of information used is that mixed $[L,\ell]$ commutators close on $\ell$ operators, and total weight added is still a well defined concept on which one can base an inductive proof, up to the fact that we had to multiply this by $n_{\rm lcm}$ to make it an integer.

Thus, we are left exploring terms of the form
\begin{align}
&\Bigg\langle \Bigg(
 \Bigg[
\left(\prod_{i_{1,1}=1}^{i_{1,1,\rm max}}\ell_{k_{1,1,i_{1,1}}/n_{1,1}}\right)\underline{\sigma}_{(1,2,\cdots,n_{1,1})} \nonumber \\
&
\qquad \;\; \left(\prod_{i_{1,2}=1}^{i_{1,2,\rm max}}\ell_{k_{1,2,i_{1,2}}/n_{1,2}}\right)\underline{\sigma}_{(n_{1,1}+1, n_{1,1}+2, \cdots,n_{1,1}+n_{1,2})} \cdots\Bigg]\Bigg)(z_1) \times \mbox{Other Insertions} \Bigg\rangle \label{noBigL}
\end{align}
where the ``Other Insertions'' are of the same form, with no full Virasoro excitations acting.

Now, we see that the above computations always lift to correlators involving covering space Virasoro excitations at ramified points only.  This suggests constructing a proof based on the covering space dimension, rather than the base space dimension.  We define the following integer
\begin{equation}
K_{\uparrow, {\rm tot,max}}=-\sum_{q} \sum_{j_q} \sum_{i_{q,j_q}} k_{q,j_q,i_{q,j_q}}
\end{equation}
which effectively multiplies the added-dimension of each product of $\ell$'s by the size of the cycle $n_{q,j_q}$ to which they are attached.  When lifting to the covering surface, this does not tell you the covering space dimension-added of each term in the computation, but rather the term with the {\it maximum} dimension-added that appears in the computation, which will appear with a sum of lower dimension-added terms as well.  This is the integer that we will use to construct an inductive proof of the following statement.

We now wish to show that all correlators of the form (\ref{noBigL}) may be written as a sum of terms of the form
\begin{align}
&\Bigg\langle \Bigg(
 \Bigg[(\ell_{-1/n_{1,1}})^{p_{1,1}}\underline{\sigma}_{(1,2,\cdots,n_{1,1})} \nonumber \\
&
\qquad \;\; (\ell_{-1/n_{1,2}})^{p_{1,2}}\underline{\sigma}_{(n_{1,1}+1, n_{1,1}+2, \cdots,n_{1,1}+n_{1,2})} \cdots\Bigg]\Bigg)(z_1) \times \mbox{Other Insertions} \Bigg\rangle \label{noBigLsimple}
\end{align}
where the other insertions are of the exact same form, where the {\it only} the excitations that appear are powers of $\ell_{-1/n_{q,j_q}}$ acting on each cycle.

Again, we start with the inductive assumption that we have already shown that all correlators of the form (\ref{noBigL}) with $K_{\uparrow, {\rm tot,max}}\leq K_0$ may be written in the form (\ref{noBigLsimple}).  We now rely on the fact that one can write a basis of terms with all $k_{q,j_q,i_{q,j_q}}\leq-1$, i.e. they are excitations, but also such that $k_{q,j_q,i_{q,j_q}}\leq k_{q,j_q,i_{q,j_q}+1}$, i.e. the operators have been ordered with ``largest dimension to the left''.

{\bf Case 1} is when there are no $\ell_{k_{q,j_q,i_{q,j_q}}}$ operators with $k_{q,j_q,i_{q,j_q}}\leq -2$, in which case it is already of the form (\ref{noBigLsimple}).  {\bf Case 2} is when there is at least one term with $k_{q,j_q,i_{q,j_q}}\leq -2$, which we may assume, without loss of generality, is the first cycle of the first operator.  Since this cycle has at least one $k_{1,1,i_{1,1}}\leq -2$, the $\ell$ operator appearing all the way to the left must satisfy $k_{1,1,1}\leq -2$ as well (because the $\ell$'s are ordered).  Writing this out explicitly, we find
\begin{align}
&\Bigg\langle \Bigg(
 \Bigg[
\left(\ell_{k_{1,1,1}}\prod_{i_{1,1}=2}^{i_{1,1,\rm max}}\ell_{k_{1,1,i_{1,1}}/n_{1,1}}\right)\underline{\sigma}_{(1,2,\cdots,n_{1,1})} \nonumber \\
&
\qquad \;\; \left(\prod_{i_{1,2}=1}^{i_{1,2,\rm max}}\ell_{k_{1,2,i_{1,2}}/n_{1,2}}\right)\underline{\sigma}_{(n_{1,1}+1, n_{1,1}+2, \cdots,n_{1,1}+n_{1,2})} \cdots\Bigg]\Bigg)(z_1) \times \mbox{Other Insertions} \Bigg\rangle \nonumber \\
&=\Bigg\langle \Bigg(
 \Bigg[
\oint \frac{dz}{2\pi i} \left(z-z_1\right)^{k_{1,1,1}/n_{1,1}+1}T_{\omega^{-k_{1,1,1}}}(z)\left(\prod_{i_{1,1}=2}^{i_{1,1,\rm max}}\ell_{k_{1,1,i_{1,1}}/n_{1,1}}\right)\underline{\sigma}_{(1,2,\cdots,n_{1,1})} \nonumber \\
&
\qquad \;\; \left(\prod_{i_{1,2}=1}^{i_{1,2,\rm max}}\ell_{k_{1,2,i_{1,2}}/n_{1,2}}\right)\underline{\sigma}_{(n_{1,1}+1, n_{1,1}+2, \cdots,n_{1,1}+n_{1,2})} \cdots\Bigg]\Bigg)(z_1) \times \mbox{Other Insertions} \Bigg\rangle. \label{noBigLsimpleCont}
\end{align}
The obstruction encountered elsewhere \cite{Burrington:2018upk, DeBeer:2019oxm} now becomes clear.  The contour measure $\left(z-z_1\right)^{k_{1,1,1}/n_{1,1}+1}$ is only adapted to the first cycle in the operator at $z_1$.  Pulling this contour would have the following obstructions.  The point $z=\infty$ is a branch point of the factor $\left(z-z_1\right)^{k_{1,1,1}/n_{1,1}+1}$ but $T_{\omega^{-k_{1,1,1}}}(z)$ is single valued, and so a contour around this point is not single valued.  Further, the cycles at other operator insertions $z\neq z_1$ can involve copies that were used to construct $T_{\omega^{-k_{1,1,1}}}(z_1)$, which causes $T_{\omega^{-k_{1,1,1}}}(z_1)$ to be non-single valued at these other operator insertions as well, even though $\left(z-z_1\right)^{k_{1,1,1}/n_{1,1}+1}$ is single valued, causing the integral measure to be non-single valued at some of the other operator insertions as well.  Thus, the contour pull is ``caught'' and one would have to integrate along the branch cuts as well.

We avoid these complications by lifting to the covering surface and examining the computation there.  The lifted computation becomes
\begin{align}
&\rightarrow \left\langle \oint_{t_{1,1}} \frac{dt}{2\pi i} \frac{\left(z(t)-z_1\right)^{k_{1,1,1}/n_{1,1}+1}}{\pa z(t)} \left(T(t)-\frac{c}{12}\left\{z(t),t\right\}\right) \phi_{1,1}'(t_1) \times \prod_{\{q,j_q\}/\{1,1\}} \phi'_{q,j_q}(t_q,j_q) \right\rangle \label{ointFormFract}
\end{align}
where $\phi'_{1,1}(t_1)$ is the lift of the excited cycle operator $\left(\prod_{i_{1,1}=2}^{i_{1,1,\rm max}}\ell_{k_{1,1,i_{1,1}}/n_{1,1}}\right)\underline{\sigma}_{(1,2,\cdots,n_{1,1})}$ appearing in (\ref{noBigLsimpleCont}).  The other $\phi'_{q,j_q}$ are the lifts of excited operators appearing at other cycles and operators in (\ref{noBigLsimpleCont}).

In the above expression, again one cannot simply pull the contour.  Even though $T(t)$ is single valued on the entire $t$ plane, the part of the measure $(z(t)-z_1)^{k_{1,1,1}/n_{1,1}+1}$ is not.  It has branch points wherever the fuction $z(t)=z_1$, which happens at some ramified and some un-ramified points in the map $z(t)$, but also at points where $z(t)\rightarrow \infty$.  The function $z(t)$ is adapted to eliminate the branch cut only at the ramified point $t_{1,1}$ appearing in $(z(t)-z_1)^{k_{1,1,1}/n_{1,1}+1}$.  Furthermore, the ramified points pick up contributions from the Schwarzian term as well, thus, at first glance, this contour seems equally obstructed.  We have eliminated the problem of non-single-valuedness of $T_{\omega^{-k_{1,1,1}}}(z)$ around other operators, while the non-single-valuedness of $(z(t)-z_1)^{k_{1,1,1}/n_{1,1}+1}$ around points where $z(t)=z_1, \infty$ remain.  There may be a base space understanding of this the elimination of the non-single-valuedness of $T_{\omega^{-k_{1,1,1}}}(z)$, although we will not discuss this at length here. \footnote{Removing this concern may be possible by explicitly examining the contour pull from the coverings space point of view.  The deformation of the contour on the cover is not constrained to have the images directly overlap in the base space.  Rather, this contour pull on the cover may dissociate various copies $T_a$ appearing in $T_{\omega^{-k_{1,1,1}}}(z)$ along separate contours and transport these dissociated contours through branch cuts, changing the copy indices $a$ appearing in the definition of $T_{\omega^{-k_{1,1,1}}}(z)$.  This seems like a natural way to construct this trade-off in a way that makes sense on the base space.}

However, the discussion of the last few sections now makes it clear how to overcome this obstacle.  The contour integral appearing above is really a power series expansion of $\mL$ operators, where the powers appearing are all integers because $z(t)=z_1+{\mc O}\left((t-t_{1,1})^{n_{1,1}}\right)$ is adapted to the point $t=t_{1,1}$ to give the proper ramification.  Furthermore, this power series has a term $\mL_{k_{1,1,1}}$ where the index is most negative.  However, this is a {\it finite} power series, because the lifted operator $\phi'_{1,1}(t_{1,1})$ has a lowest conformal dimension ancestor. Thus, the contour integral in \ref{ointFormFract} may be expanded as
\begin{align}
&\rightarrow \left\langle \left(\sum_{j_{1,1,1}=k_{1,1,1}}^{-2} a_{j_{1,1,1}}\mL_{j_{1,1,1}} +a_{0,S}+ \sum_{j_{1,1,1}=-1}^{j_{1,1,1,{\rm max}}} a_{j_{1,1,1}}\mL_{j_{1,1,1}}\right)\phi'_{1,1}(t_{1,1}) \times \prod_{\{q,j_q\}/\{1,1\}} \phi'_{q,j_q}(t_q,j_q) \right\rangle
\end{align}
Recall the meaning of the indices $j_{q,j_q,i_{q,j_q}}$ to properly label the subscripts of the $\mL$'s: $j_{q,j_q,i_{q,j_q}}$ refers to the $j_{q,j_q,i_{q,j_q}}$-th term in the summation expansion of the $\ell$ operator expressed on the cover, each $\ell$ is itself labeled as the $i_{q,j_q}$-th $\ell$ operator appearing in the product of $\ell$'s, which acts on the $j_{q}$-th cycle of the $q$-th operator (group element).  Thus, simply by the name $j_{1,1,1}$ we know we are looking at the first operator, first cycle, first $\ell$ in the product of $\ell$'s, to which each of these indices refer.  Going to different operators changes the first index, going to different cycle within the same operator changes the second index, and going to a different member of the product of $\ell$'s at a given operator at a given cycle changes the third index.  The coefficients $a_{j_{q,j_q,i_{q,j_q}}}$, which must be labeled the same way, contain map information when expanding the term $(z(t)-z_1)^{k_{1,1,1}/n_{1,1}+1}/\pa z(t)$ to the proper order.  In addition, the extra term $a_{0,S}$ is the term coming from the Schwarzian, which is a fixed function, and may be easily absorbed into the ${\mc L}_0$ coefficient, given that the operator to the right has definite scaling dimension.

Now the conclusion is obvious.  The above is a finite sum of terms of excitations $\mL$ acting on some operator.  The basic theorem in the introduction now applies to the computation on the covering surface, i.e. that the above correlator on the covering surface may be written in terms of the ancestor correlator and {\it covering space} derivatives.  These covering space derivatives may be replaced with $\mL_{-1}$'s which may be found in terms of $\ell_{-1/n_{q,j_q}}$'s using the weight reconstruction.  Lower dimension operators will also appear in these give only $\ell_0$'s and operators which may be removed by commutation relations.  Hence, the form (\ref{noBigLsimple}) is achieved because the $\ell_0$'s may be replaced with their eigenvalues.

While the correlators on the covering surface may be reduced in this way, the functions that result will be most naturally written in terms of base space data {\it and} covering space data.  While it is not obvious how, we expect that these combine into expressions that make sense in the base space.  The direct proof of such a statement presumably depends on the fact that the contour integrals describing the excitations are initially defined on the base space, with the covering space playing an auxiliary (although very useful) role.  We explicitly calculate some examples in the next section to better test this expectation.

A very important case comes up frequently, which simplifies matters drastically: the case of bare twists as the ancestors.  Bare twists are those operators that lift to the identity on the cover.  In such a situation, $\mL_{-1} \cdot 1=0$, and so none of the extra complications arise.   The only term in the sum over terms of the form (\ref{noBigLsimple}) are those that contain no powers of $\ell_{-1/n_{q,i}}$ at all, i.e. just the original correlator
\begin{align}
&\Bigg\langle \Bigg({\sigma}_{(n_{1,1}+1, n_{1,1}+2, \cdots,n_{1,1}+n_{1,2})} \cdots\Bigg)(z_1) \times \mbox{Other Bare Twists} \Bigg\rangle. \label{BareOnly}
\end{align}
Above, we have used the notation $\sigma$ without any extra marks to indicate a bare twist which lifts to an identity operator on the cover.  In such a situation, we have shown that the form of the general correlator (\ref{bigLpullBase}) after replacing $\underline{\sigma}\rightarrow  \sigma$ becomes
\begin{align}
\left(\mbox{Algorithmic Map Data}\right)\times \Bigg\langle \Bigg({\sigma}_{(n_{1,1}+1, n_{1,1}+2, \cdots,n_{1,1}+n_{1,2})} \cdots\Bigg)(z_1) \times \mbox{Other Bare Twists} \Bigg\rangle
\end{align}
and {\it base space} derivatives of such terms (if full Virasoro excitations were part of the original correlator).  Above, the ``Algorighmic Map Data'' comes from repeated expansions of $z(t)$ at different points in the cover, and so can be done in general, given the map $z(t)$ (finding $z(t)$ is, of course, a non-trivial step).  From the base space point of view, $\ell_{-1/n}$ annihilate the bare twists, i.e. they are null vectors in the tower of excitations.   One may use this fact to calculate certain classes of correlators \cite{Dei:2019iym}.  Thus the conclusions for correlators in this subsector can be written entirely in terms of the correlators of bare twists, and {\it base space} derivatives.  Furthermore, from our previous work \cite{Burrington:2018upk}, it appears that this is a closed subsector in the large $N$ limit.

Again, we emphasize that because the above was written in terms of operators understandable from the base space (\ref{bigLpullBase}), we expect that the result must simplify to an expression that makes sense on the base-space: the ``Algorighmic Map Data'' must combine into base-space data in some (possibly complicated) way, which we explore through examples in the next section.

\section{Examples for (4)-(2)-(5) fusion}

While the proofs in the previous sections seem quite detailed, the lessons are fairly clear and straightforward (although lengthy to implement).  When one sees a full Virasoro mode with a subscript $-2$ or less, one pulls the contour on the base space, leading to a sum of lower dimension-added correlators.  Any $L_{m}$ with $m>0$ may be replaced using commutation relations, and because the mixed commutation relations $[L,\ell]$ close on $\ell$ will result in a set of lower dimension-added correlators which maintain the order with the $L$ operators to the left.  One continues this method until the only full Virasoro generators remaining are $L_{-1}$'s and $L_0$'s operating at various points, which are derivatives and eigenvalues.  This leaves us to contend with the fractional Virasoro modes.  We see that the fractional Virasoro modes with index $-2/(n_{q,j_q})$ or less may be pulled on the cover, arriving at a sum of correlators written in terms of less total added-dimension on the cover, eventually leading to a correlator with only $\mL_{-1}$ and $\mL_{0}$ on the cover, which may be replaced with covering space derivatives and eigenvalues.  Any Virasoro mode with subscript greater than 0 cannot be pulled due to an obstruction at $z=\infty$ or $t=\infty$, but it is also undesirable to try to pull the contour for such a mode: the mode is either an eigenvalue, or lowers the dimension-added for the operator in question, and so commutation relations can be used to remove it from discussion.

The arguments of the last section reduce excited correlators to the correlators of primary on the cover, and one may be curious as to how to obtain correlators of primaries on the cover directly from the CFT at hand.  We breifly address this here.  Given a primary operator in the seed CFT, one can construct the fractional excitations in the orbifold by taking linear combinations with relative phases, just as with the stress tensor.  This allows the construction of fractional modes in the vicinity of an $n_i$ cycle, and one such mode is $\mathcal{O}_{-a_i/n_i}$ where $a_i$ is the weight of the original operator in the base space, and the subscript denotes that we are taking a particular mode at this point. When acting on a bare twist, this mode lifts to the cover as $\oint \frac{dt}{2\pi i}(z(t)-z_n)^{-a_i/n_i + a_i -1}(\pa z)^{-a_i+1} \mathcal{O}_{a_i}(t)$ (these terms are simpler than the expressions for the stress tensor: the stress tensor is not primary).  Seeing that $z(t)-z_{n_i}$ behaves as $(t-t_{n_i})^{n_i}$ on the cover, the leading order term in the prefactor behaves as $1/(t-t_{n_i})$, and so gives just the primary $\mathcal{O}_{a_i}(t_{n_i})$ insertion on the cover (dressed with some map data).  In fact, given the expansion for the map to the covering space
\begin{equation}
(z-z_i)=A_i (t-t_i)^{n_i}+\cdots
\end{equation}
one can generally write that the operator
\begin{equation}
\mathcal{O}_{-a_i/n_i} \sigma(z_i) \rightarrow A_i^{-a_i/n_i} \frac{1}{(n_i)^{a_i-1}} \mathcal{O}_{a_i}(t_{n_i}) \label{primarylift}
\end{equation}
where the above unadorned $\sigma$ indicates a bare twist which lifts to the identity on the cover.  One can also argue that this is the correct mode of the operator by examining the eigenvalue that is produced by operating $\ell_0$ on an object that lifts to a primary on the cover: the weight added to the bare-twist weight is just $a_i/n_i$, suggesting the mode $\mathcal{O}_{-a_i/n_i}$.

We now turn to computing some example $n$-point functions, for which we concentrate on the 3-point correlator of single cycle twists.  These have several features which are particularly helpful.  First, the maps are known \cite{Lunin:2000yv} for specified locations of the twists and ramified points in the map.  Furthermore, because these maps are 3-point functions on the base and there are only 3 ramified points in the map from the cover, we see that the $SL(2)$ symmetries on the base and on the cover will be sufficient to fully parameterize the cover 3-point functions.  This makes the covering space derivatives easy to compute, and distinguish from base space derivatives.  In addition, higher point correlation functions on the cover need to be treated with some care: the map data which parameterizes the cross ratio on the cover specifies both the cross ratio on the base and specifies a certain amount about which type of group product is being considered, and which crossing channel (equally, which exact group element) is being considered in the correlator \cite{Lunin:2000yv,Pakman:2009zz,Pakman:2009ab,GarciaiTormo:2018vqv,Burrington:2018upk,DeBeer:2019oxm,Keller:2019suk,Lima:2020boh,Lima:2020urq,Lima:2021wrz,AlvesLima:2022elo}.  We will consider these higher point functions in future work.

Thus, the 3 point functions of single cycle operators are the most simple to analyze.  We further specify to a specific fusion of a twist $(4)-(2)-(5)$ fusion to make all calculations easy to follow.  Although a treatment of the arbitrary single cycle 3-point functions can be found (the maps are known \cite{Lunin:2000yv}), we leave this more involved calculation for a future publication.

First, consider the map \cite{Lunin:2000yv}
\begin{equation}
w=-u^4(-5+4u)
\end{equation}
which has ramified points at $u=0,1,\infty$ with single cycle twists of order $4,2,5$ respectively; this may be confirmed by expanding the above function around these points.  However, we wish to map both the base space problem and the covering space problem to finite points to allow the evaluation of derivatives at these points.  Thus, we use successive $SL(2)$ transformations
\begin{align}
&w=\frac{(z_1-z_\infty)}{(z_1-z_0}\frac{(z-z_0)}{(z-z_\infty)} \nn \\
&u=\frac{(t_1-t_\infty)}{(t_1-t_0)}\frac{(t-t_0)}{(t-t_\infty)}
\end{align}
These effectively map the points of interest on the $u$-cover $w=0\rightarrow z=z_0, w=1 \rightarrow z=z_1, w=\infty \rightarrow z=z_\infty$, and similarly for the $t$ coordinate.  This allows us to write the composite map as
\begin{equation}
z(t)=\frac{t_{1,0}^5(t-t_\infty)^5z_0z_{1,\infty}-5t_{1,0}t_{1,\infty}^4(t-t_0)^4(t-t_\infty)z_{1,0}z_\infty +4t_{1,\infty}^5(t-t0)^5z_{1,0}z_\infty}
 {t_{1,0}^5(t-t_\infty)^5z_1-t_{1,0}^5(t-t_\infty)^5z_\infty-5t_{1,0}t_{1,\infty}^4(t-t_0)^4(t-t_\infty)z_{1,0}+4t_{1,\infty}^5 (t-t_0)^5z_{1,0}} \label{zmap245}
\end{equation}
where above we have use the standard notation $z_{i,j}=z_i-z_j$ and $t_{i,j}=t_i-t_j$.  Again, one may simply expand the above map near $t=t_0, t=t_1, t=t_\infty$ to verify that the map is correctly ramified for a twist $(4)-(2)-(5)$ fusion.

We will need several parts from the above equation.  We first compute the Schwarzian, and find
\begin{align}
&\left\{z,t\right\}=-\frac{24}{2(t-t_\infty)^2}-\frac{15}{2(t-t_0)^2}-\frac{3}{2(t-t_1)^2} \nn \\
&\qquad \qquad \qquad +\frac{18}{(t-t_\infty)(t-t_0)}+\frac{6}{(t-t_\infty)(t-t_1)}+\frac{-3}{(t-t_1)(t-t_0)}
\end{align}
Next we consider the expansions about the ramified points:
\begin{align}
& (z-z_0)=A_0(t-t_0)^4(1+S_{1,0}(t-t_0)^1+S_{2,0}(t-t_0)^2+\cdots) \nn \\
& (z-z_1)=A_1(t-t_0)^2(1+S_{1,1}(t-t_1)^1+S_{2,1}(t-t_1)^2+\cdots) \\
& (z-z_\infty)=A_\infty (t-t_\infty)^5(1+S_{1,\infty}(t-t_\infty)^1+S_{2,\infty}(t-t_\infty)^2+\cdots) \nn
\end{align}
with
\begin{align}
& A_0=5\frac{z_{\infty,0}z_{1,0}}{z_{\infty,1}}\left(\frac{t_{\infty,1}}{t_{\infty,0}t_{1,0}}\right)^4 \nn \\
& A_1=-10\frac{z_{\infty,1}z_{1,0}}{z_{\infty,0}}\left(\frac{t_{\infty,0}}{t_{\infty,1}t_{1,0}}\right)^2 \\
& A_\infty=-\frac{1}{4}\frac{z_{\infty,1}z_{\infty,0}}{z_{1,0}}\left(\frac{t_{1,0}}{t_{\infty,1}t_{\infty,0}}\right)^5 \nn
\end{align}
and further terms parameterized by the $S_{"{\rm order}","{\rm location}"}$ polynomials in $t_i,z_i$, which are straightforward to compute from (\ref{zmap245}).

We now start with a base correlator, constructed such that the lift to the cover becomes a correlator of primaries.  We start with primary fields $\mathcal{O}_{\infty}, \mathcal{O}_1, \mathcal{O}_0$ with weights $a_\infty, a_1, a_0$ respectively.  We use (\ref{primarylift}), and find
\begin{align}
&\langle \left[\mathcal{O}_{\infty,-a_\infty/5} \sigma_{(5)}\right](z_\infty)\left[\mathcal{O}_{1,-a_1/2} \sigma_{(2)}\right](z_1)\left[\mathcal{O}_{0,-a_0/4} \sigma_{(4)}\right](z_0)\rangle \nn \\
&\rightarrow A_\infty^{-a_\infty/5} \frac{1}{(5)^{a_\infty-1}} A_1^{-a_1/2} \frac{1}{(2)^{a_1-1}} A_0^{-a_0/4} \frac{1}{(4)^{a_0-1}} \langle \mathcal{O}_\infty (t_\infty) \mathcal{O}_1(t_1)\mathcal{O}_0(t_0)\rangle\nn \\
& =\frac{(5)^{-1/4}(-10)^{-1/2}(-1/4)^{-1/5} }{(5)^{a_\infty-1}(2)^{a_1-1}(4)^{a_0-1}}\left(\frac{z_{\infty,0}z_{1,0}}{z_{\infty,1}}\right)^{-a_0/4} \left(\frac{z_{\infty,1}z_{1,0}}{z_{\infty,0}}\right)^{-a_1/2}\left(\frac{z_{\infty,1}z_{\infty,0}}{z_{1,0}}\right)^{-a_\infty/5} \nn \\
& \times
\left(\frac{t_{\infty,1}}{t_{\infty,0}t_{1,0}}\right)^{-a_0}
\left(\frac{t_{\infty,0}}{t_{\infty,1}t_{1,0}}\right)^{-a_1}
\left(\frac{t_{1,0}}{t_{\infty,1}t_{\infty,0}}\right)^{-a_\infty}
\langle \mathcal{O}_\infty (t_\infty) \mathcal{O}_1(t_1)\mathcal{O}_0(t_0)\rangle
\end{align}
We see that the above prefactor of $t_{i,j}$ are just so as to cancel the factors of $t_{i,j}$ in the three point function $\langle \mathcal{O}_\infty (t_\infty) \mathcal{O}_1(t_1)\mathcal{O}_0(t_0)\rangle$, and further that the powers of $z_{i,j}$ are just so as to combine with the Liouville term from \cite{Lunin:2000yv} to shift the weights in the base three point function.  Thus, we have that these combine into a total three point function on the base
\begin{align}
&\langle \left[\mathcal{O}_{\infty,-a_\infty/5} \sigma_{(5)}\right](z_\infty)\left[\mathcal{O}_{1,-a_1/2} \sigma_{(2)}\right](z_1)\left[\mathcal{O}_{0,-a_0/4} \sigma_{(4)}\right](z_0)\rangle= \nn\\
&\frac{(5)^{-1/4}(-10)^{-1/2}(-1/4)^{-1/5} }{(5)^{a_\infty-1}(2)^{a_1-1}(4)^{a_0-1}} C_{4,2,5} C_{a_0,a_1,a_\infty} \left(\frac{z_{\infty,0}z_{1,0}}{z_{\infty,1}}\right)^{h_{0,{\rm tot}}}
\left(\frac{z_{\infty,1}z_{1,0}}{z_{\infty,0}}\right)^{h_{1,{\rm tot}}} \left(\frac{z_{\infty,1}z_{\infty,0}}{z_{1,0}}\right)^{h_{\infty,{\rm tot}}}
\end{align}
where $C_{4,2,5}$ is the structure constant computed in \cite{Lunin:2000yv} for the bare twists, and $C_{a_0,a_1,a_\infty}$ is the structure constant from the unspecified operators on the cover.  The total weights are given as
\begin{equation}
h_{0,{\rm tot}}=\frac{c}{24}(4-1/4) + \frac{a_0}{4}, \qquad h_{1,{\rm tot}}=\frac{c}{24}(2-1/2) + \frac{a_1}{2}, \qquad h_{\infty,{\rm tot}}=\frac{c}{24}(5-1/5) + \frac{a_\infty}{5}.
\end{equation}
The ambiguity in the choice of phase in the $n-th$ roots above may be removed by considering mirrored excitations in the right moving sector.  Henceforth, we simply combine the above information into an overall structure constant for the correlator
\begin{align}
&\langle \left[\mathcal{O}_{\infty,-a_\infty/5} \sigma_{(5)}\right](z_\infty)\left[\mathcal{O}_{1,-a_1/2} \sigma_{(2)}\right](z_1)\left[\mathcal{O}_{0,-a_0/4} \sigma_{(4)}\right](z_0)\rangle= \nn\\
&C^{*}_{4,2,5} \left(\frac{z_{\infty,0}z_{1,0}}{z_{\infty,1}}\right)^{h_{0,{\rm tot}}}
\left(\frac{z_{\infty,1}z_{1,0}}{z_{\infty,0}}\right)^{h_{1,{\rm tot}}} \left(\frac{z_{\infty,1}z_{\infty,0}}{z_{1,0}}\right)^{h_{\infty,{\rm tot}}}.
\end{align}
It is this base correlator which we will write all subsequent correlators in terms of, the other correlators being constructed from operators excited by fractional Virasoro modes.

We begin by exploring the lowest level excitations acting at each position.  As one possible example, we wish to compute
\begin{align}
&\frac{\langle \left[\ell_{-1/5}\mathcal{O}_{\infty,-a_\infty/5} \sigma_{(5)}\right](z_\infty)\left[\mathcal{O}_{1,-a_1/2} \sigma_{(2)}\right](z_1)\left[\mathcal{O}_{0,-a_0/4} \sigma_{(4)}\right](z_0)\rangle}{\langle \left[\mathcal{O}_{\infty,-a_\infty/5} \sigma_{(5)}\right](z_\infty)\left[\mathcal{O}_{1,-a_1/2} \sigma_{(2)}\right](z_1)\left[\mathcal{O}_{0,-a_0/4} \sigma_{(4)}\right](z_0)\rangle}\nn \\
&=\frac{\langle\left[\oint\frac{dt}{2\pi i}(z-z_\infty)^{-1/5+1}(\pa z)^{-1}\left(T(t)-\frac{c}{12}\left\{z,t\right\}\right)\mathcal{O}_{\infty}\right](t_{\infty}) \mathcal{O}_{1}(t_{1})\mathcal{O}_{0}(t_{0})\rangle}{\langle \mathcal{O}_\infty (t_\infty) \mathcal{O}_1(t_1)\mathcal{O}_0(t_0)\rangle}.
\end{align}
Note that in the above the $\rightarrow$ notation has not been used: the Liouville parts in the numerator and denominator cancel, so the above is an equality.  Using the above expansions, we find that the schwarzian term above does not contribute, which is to be expected: in the case of bare twists the mode $\ell_{-1/n}$ annihilates the state (such null vectors have been used to calculate certain classes of correlators \cite{Dei:2019iym}).  Thus, we are left computing the terms involving the stress tensor.  After evaluating the expansion, we find the above becomes
\begin{align}
&=\frac{\langle\left[\oint\frac{dt}{2\pi i}\frac{A_\infty^{-1/5}}{5} \left({\mc L}_{-1}-\frac{1}{2} \frac{t_0+3t_1-4t_\infty}{t_{\infty,0}t_{\infty,1}}{\mc L}_0\right)\mathcal{O}_{\infty}\right](t_{\infty}) \mathcal{O}_{1}(t_{1})\mathcal{O}_{0}(t_{0})\rangle}{\langle \mathcal{O}_\infty (t_\infty) \mathcal{O}_1(t_1)\mathcal{O}_0(t_0)\rangle}. \nn \\
\end{align}
Of course ${\mc L}_{-1}$ is a covering space derivative acting on a known functional form, and ${\mc L}_0$ is just the eigenvalue $a_\infty$.  This allows us to simplify the result, finding
\begin{align}
&\frac{\langle \left[\ell_{-1/5}\mathcal{O}_{\infty,-a_\infty/5} \sigma_{(5)}\right](z_\infty)\left[\mathcal{O}_{1,-a_1/2} \sigma_{(2)}\right](z_1)\left[\mathcal{O}_{0,-a_0/4} \sigma_{(4)}\right](z_0)\rangle}{\langle \left[\mathcal{O}_{\infty,-a_\infty/5} \sigma_{(5)}\right](z_\infty)\left[\mathcal{O}_{1,-a_1/2} \sigma_{(2)}\right](z_1)\left[\mathcal{O}_{0,-a_0/4} \sigma_{(4)}\right](z_0)\rangle}\nn \\
&= \frac{A_\infty^{-1/5}}{5}\frac{t_{1,0}}{t_{\infty,0}t_{\infty,1}}\left(a_0-a_1-\frac{a_\infty}{2}\right) \nn \\
&=\frac{1}{5}\left(-\frac{1}{4}\right)^{-1/5}\left(\frac{z_{\infty,1}z_{\infty,0}}{z_{1,0}}\right)^{-1/5} \left(a_0-a_1-\frac{a_\infty}{2}\right)
\end{align}
We see that the right hand side no longer contains covering space information ($t_{i,j}$), and thus the descendant correlator has been written in terms of the ancestor, as desired.  Furthermore, the factors of $z_{ij}$ are just so as to modify the $z_{ij}$ dependence so that the new correlator is also the form of a correlator of primaries.  This is simply because the excitation $\ell_{-1/5}$ has insufficient weight to prevent $L_{m}$ with $m>0$ from annihilating this excitation.

Similar calculations give
\begin{align}
&\frac{\langle \left[\mathcal{O}_{\infty,-a_\infty/5} \sigma_{(5)}\right](z_\infty)\left[\ell_{-1/2}\mathcal{O}_{1,-a_1/2} \sigma_{(2)}\right](z_1)\left[\mathcal{O}_{0,-a_0/4} \sigma_{(4)}\right](z_0)\rangle}{\langle \left[\mathcal{O}_{\infty,-a_\infty/5} \sigma_{(5)}\right](z_\infty)\left[\mathcal{O}_{1,-a_1/2} \sigma_{(2)}\right](z_1)\left[\mathcal{O}_{0,-a_0/4} \sigma_{(4)}\right](z_0)\rangle}\nn \\
&=\frac{1}{2}\left(-10\right)^{-1/2}\left(\frac{z_{\infty,1}z_{1,0}}{z_{\infty,0}}\right)^{-1/2} \left(a_0-a_\infty+3a_1\right) \\
&\frac{\langle \left[\mathcal{O}_{\infty,-a_\infty/5} \sigma_{(5)}\right](z_\infty)\left[\mathcal{O}_{1,-a_1/2} \sigma_{(2)}\right](z_1)\left[\ell_{-1/4}\mathcal{O}_{0,-a_0/4} \sigma_{(4)}\right](z_0)\rangle}{\langle \left[\mathcal{O}_{\infty,-a_\infty/5} \sigma_{(5)}\right](z_\infty)\left[\mathcal{O}_{1,-a_1/2} \sigma_{(2)}\right](z_1)\left[\mathcal{O}_{0,-a_0/4} \sigma_{(4)}\right](z_0)\rangle}\nn \\
&=\frac{1}{4}\left(5\right)^{-1/4}\left(\frac{z_{\infty,0}z_{1,0}}{z_{\infty,1}}\right)^{-1/4} \left(\frac{-1}{5}\left(5a_1-5a_\infty +7a_0\right)\right)
\end{align}
As a check, one may consider the case where the excitation acts on a bare twist.  In such a situation, the $a_i$ at the location of the excitation is 0, and the $a_i$ at the other locations must be equal (the lifted correlator is a two point function on the cover).  This renders the right hand side $0$, as it must be: the mode $\ell_{-1/n}$ acting on a bare twist gives 0, completing the check.

We now move to an example where a contour pull will be advantageous.  We consider the excitation
\begin{align}
&\frac{\langle \left[\ell_{-2/5}\mathcal{O}_{\infty,-a_\infty/5} \sigma_{(5)}\right](z_\infty)\left[\mathcal{O}_{1,-a_1/2} \sigma_{(2)}\right](z_1)\left[\mathcal{O}_{0,-a_0/4} \sigma_{(4)}\right](z_0)\rangle}{\langle \left[\mathcal{O}_{\infty,-a_\infty/5} \sigma_{(5)}\right](z_\infty)\left[\mathcal{O}_{1,-a_1/2} \sigma_{(2)}\right](z_1)\left[\mathcal{O}_{0,-a_0/4} \sigma_{(4)}\right](z_0)\rangle}\nn \\
&=\frac{\langle\left[\oint\frac{dt}{2\pi i}(z-z_\infty)^{-2/5+1}(\pa z)^{-1}\left(T(t)-\frac{c}{12}\left\{z,t\right\}\right)\mathcal{O}_{\infty}\right](t_{\infty}) \mathcal{O}_{1}(t_{1})\mathcal{O}_{0}(t_{0})\rangle}{\langle \mathcal{O}_\infty (t_\infty) \mathcal{O}_1(t_1)\mathcal{O}_0(t_0)\rangle} \nn \\
&=\frac{A_\infty^{-2/5}}{5}\frac{\langle\left[\left({\mc L}_{-2}+F_{2,1,\infty}{\mc L}_{-1} +F_{2,0,\infty}{\mc L}_{0}-\frac{c}{16}\frac{t_{1,0}^2}{t_{\infty,1}^2t_{\infty,0}^2}\right)\mathcal{O}_{\infty}\right](t_{\infty}) \mathcal{O}_{1}(t_{1})\mathcal{O}_{0}(t_{0})\rangle}{\langle \mathcal{O}_\infty (t_\infty) \mathcal{O}_1(t_1)\mathcal{O}_0(t_0)\rangle}.
\end{align}
One may now pull the contour defining ${\mc L}_{-2}$, resulting in a sum of operators that act at the other points on the covering surface with operator insertions.  However, this has been done in generality when there are only operators exciting on operator in the correlator: see \cite{DiFrancesco:1997nk} equation (6.152).  Therefore, we may simply replace the above ${\mc L}_i$ operators with the differential operators of \cite{DiFrancesco:1997nk}.  This leads to
\begin{align}
&\frac{\langle \left[\ell_{-2/5}\mathcal{O}_{\infty,-a_\infty/5} \sigma_{(5)}\right](z_\infty)\left[\mathcal{O}_{1,-a_1/2} \sigma_{(2)}\right](z_1)\left[\mathcal{O}_{0,-a_0/4} \sigma_{(4)}\right](z_0)\rangle}{\langle \left[\mathcal{O}_{\infty,-a_\infty/5} \sigma_{(5)}\right](z_\infty)\left[\mathcal{O}_{1,-a_1/2} \sigma_{(2)}\right](z_1)\left[\mathcal{O}_{0,-a_0/4} \sigma_{(4)}\right](z_0)\rangle}\nn \\
&= \frac{A_\infty^{-2/5}}{5} \frac{t_{1,0}^2}{t_{\infty,1}^2t_{\infty,0}^2}\left(\frac{-1}{16} a_\infty -\frac{1}{4} a_0 +\frac{5}{4}a_1-\frac{c}{16}\right) \nn \\
&= \frac{\left(-1/4\right)^{-2/5}\left(\frac{z_{\infty,1}z_{\infty,0}}{z_{1,0}}\right)^{-2/5}}{5} \left(\frac{-1}{16} a_\infty -\frac{1}{4} a_0 +\frac{5}{4}a_1-\frac{c}{16}\right)
\end{align}
so again, the descendant has been written in terms of the ancestor without reference to the covering space coordinates $t_i$.

We now turn to an example which more directly displays one of the major points of the last section, i.e. that the operator valued Laurent expansion truncates, and this is what allows us to complete calculations on the covering surface.  We consider the following correlator
\begin{align}
&\frac{\langle \left[\ell_{-2/5}\ell_{-5/5}\mathcal{O}_{\infty,-a_\infty/5} \sigma_{(5)}\right](z_\infty)\left[\mathcal{O}_{1,-a_1/2} \sigma_{(2)}\right](z_1)\left[\mathcal{O}_{0,-a_0/4} \sigma_{(4)}\right](z_0)\rangle}{\langle \left[\mathcal{O}_{\infty,-a_\infty/5} \sigma_{(5)}\right](z_\infty)\left[\mathcal{O}_{1,-a_1/2} \sigma_{(2)}\right](z_1)\left[\mathcal{O}_{0,-a_0/4} \sigma_{(4)}\right](z_0)\rangle}.
\end{align}
In the above we see that there are two expansions in the lift, coming from the two $\ell$ operators:
\begin{align}
&\ell_{-2/5} \rightarrow \oint \frac{dt}{2\pi i} (z-z_\infty)^{\frac{-2}{5}+1} (\pa z)^{-1} \left( T(t) -\frac{c}{12} \left\{z,t\right\}\right)  \\
&\ell_{-5/5} \rightarrow \oint \frac{dt}{2\pi i} (z-z_\infty)^{\frac{-5}{5}+1} (\pa z)^{-1} \left( T(t) -\frac{c}{12} \left\{z,t\right\}\right).
\end{align}
To explore this further, we look at the relevant expansions in the measures
\begin{align}
(z-z_\infty)^{\frac{-2}{5}+1}(\pa z)^{-1} = \frac{A_\infty^{-2/5}}{5} \frac{1}{(t-t_\infty)}\left(1+\sum_{r=-1}^\infty F_{\infty,-2,r}(t-t_\infty)^{r+2}\right) \nn \\
(z-z_\infty)^{\frac{-5}{5}+1}(\pa z)^{-1} = \frac{A_\infty^{-5/5}}{5} \frac{1}{(t-t_\infty)^4}\left(1+\sum_{r=-4}^\infty F_{\infty,-5,r}(t-t_\infty)^{r+5}\right) \label{expansionsform2m5}
\end{align}
where the coefficients $F_{"{\rm location}","{\rm index}", "{\rm order}"}$ are polynomials of $z_i,t_i$.  We have shifted the summation indices $r$ to make them count the total power of $(t-t_\infty)$ plus 1: this convention will become clear momentarily.  These are contour integrated against $T(t)$, or the schwarzian term, and then this mode is applied to an operator.  The infinite sums may therefore be truncated at a finite value: otherwise the annihilation modes become too high index.  In the sum resulting from the lift of $\ell_{-5/5}$, which is the operator applied first, this means truncating the power to maximum power $r=0$ corresponding to a mode ${\mc L}_0$ (making the shifts in the index clear).  However, second sum for $\ell_{-2/5}$ can contain annihilation operators up to $r=5$, i.e. ${\mc L}_5$, because the operator to the right contains ${\mc L}_{-5}$ as the highest dimension excitation.  Thus, we may expand the lift as
\begin{align}
& \ell_{-2/5}\ell_{-5/5} \rightarrow \nn \\
& \frac{A_{\infty}^{-7/5}}{5^2}\Big({\mc L}_{-2}+F_{\infty,-2,-1}{\mc L}_{-1}+F_{\infty,-2,0}{\mc L}_{0}+F_{\infty,-2,1}{\mc L}_{1}+F_{\infty,-2,2}{\mc L}_{2}+F_{\infty,-2,3}{\mc L}_{3}\nn \\
& \qquad \qquad \qquad \qquad \qquad \qquad \qquad \qquad \qquad \qquad +F_{\infty,-2,4}{\mc L}_{4}+F_{\infty,-2,5}{\mc L}_{5}-\frac{c}{16}\frac{t_{1,0}^2}{t_{\infty,1}^2t_{\infty,0}^2}\Big) \nn \\
&\times \Big({\mc L}_{-5}+F_{\infty,-5,-4}{\mc L}_{-4}+F_{\infty,-5,-2}{\mc L}_{-2} \nn \\
&\qquad \qquad \qquad \qquad \qquad  +F_{\infty,-5,-1}{\mc L}_{-1}+F_{\infty,-5,0}{\mc L}_{0}-\frac{c(17z_0+47z_1-64z_\infty)}{128(z_1-z_0)}\frac{t_{1,0}^5}{t_{\infty,1}^5t_{\infty,0}^5}\Big)
\end{align}
when operated on a primary on the cover.  Above, we have evaluated the terms coming from the schwarzians explicitly.  We may of course expand the above, and many terms will automatically give 0, for example ${\mc L}_5 {\mc L}_{-2}=0$ when operating on a primary.  Furthermore, terms with annihilation operators to the left of creation operators may be replaced with single operators using the Virasoro algebra on the cover, e.g. ${\mc L}_4 {\mc L}_{-5}=[{\mc L}_4, {\mc L}_{-5}]=9{\mc L}_{-1}$ (recall that on the cover there is a single copy, and so central charge terms appear as $c\times 1$).  This allows us to simplify the expression to
\begin{align}
& \ell_{-2/5}\ell_{-5/5} \rightarrow \nn \\
& \sum_{i=-2}^0 \sum_{j=-5}^0 F_{\infty,-2,i}F_{\infty,-5,j} {\mc L}_{i} {\mc L}_j +\frac{c^2}{2048}\frac{(17z_0+47z_1-64z_\infty)}{(z_1-z_0)}\frac{t_{1,0}^7}{t_{\infty,1}^7t_{\infty,0}^7} \nn \\
&-\frac{c}{16}\frac{t_{1,0}^2}{t_{\infty,1}^2t_{\infty,0}^2} \sum_{i=-5}^0F_{\infty,-5,i}{\mc L}_i  -\frac{c(17z_0+47z_1-64z_\infty)}{128(z_1-z_0)}\frac{t_{1,0}^5}{t_{\infty,1}^5t_{\infty,0}^5}\sum_{i=-2}^0F_{\infty,-2,i}{\mc L}_i\nn \\
& + \sum_{i=1}^5 \sum_{j=-5}^{-i} F_{\infty,-2,i}F_{\infty,-5,j}\left((i-j){\mc L}_{i+j} + \delta_{i+j,0}\frac{c}{12}(i^2-1)i\right)
\end{align}
when operating on a primary, and we have used the notation $F_{\infty,i,i}=1$. The expression above only contains ${\mc L}_i$ with $i\leq 0$, and furthermore this is the only site with excitations acting.  Therefore, we may use the differential operators of (6.152) in \cite{DiFrancesco:1997nk} and evaluate the answer \footnote{Correlators containing excitations acting on multiple operators may still be evaluated by pulling the contours: the use of the cited equation is simply a shortcut available in this case.}. Doing so, we find
\begin{align}
&\frac{\langle \left[\ell_{-2/5}\ell_{-5/5}\mathcal{O}_{\infty,-a_\infty/5} \sigma_{(5)}\right](z_\infty)\left[\mathcal{O}_{1,-a_1/2} \sigma_{(2)}\right](z_1)\left[\mathcal{O}_{0,-a_0/4} \sigma_{(4)}\right](z_0)\rangle}{\langle \left[\mathcal{O}_{\infty,-a_\infty/5} \sigma_{(5)}\right](z_\infty)\left[\mathcal{O}_{1,-a_1/2} \sigma_{(2)}\right](z_1)\left[\mathcal{O}_{0,-a_0/4} \sigma_{(4)}\right](z_0)\rangle} \nn \\
&=  \frac{\left(-1/4\right)^{-7/5}\left(\frac{z_{\infty,1}z_{\infty,0}}{z_{1,0}}\right)^{-7/5}}{5^2} \nn \\
& \times \Bigg(
\frac{(17z_0+47z_1-64z_\infty)c^2}{2048z_{1,0}}-\frac{(157z_0-1949z_1+1792z_\infty)c}{10240z_{1,0}} \nn \\ &\qquad +\frac{(7z_0+57z_1-64z_\infty)c a_0}{512z_{1,0}}-\frac{5(13z_0+51z_1-64z_\infty)c a_1 }{512z_{1,0}}+ \frac{(49z_0+79z_1-128z_\infty)ca_\infty}{2048z_{1,0}} \nn \\
& \qquad +\frac{5z_{1,0}a_0^2}{64}+\frac{25z_{1,0}a_1^2}{32}+\frac{(z_0+z_1-2z_\infty)a_\infty^2}{64z_{1,0}} \nn \\
& \qquad -\frac{35z_{1,0}a_0a_1}{64}+\frac{(11z_0+21z_1-32z_\infty)a_0a_\infty}{256z_{1,0}}-\frac{5(7z_0+9z_1-16z_i)a_1a_\infty}{128z_{1,0}}
\nn \\
&\qquad \qquad -\frac{(59z_0-3653z_1+3584z_\infty)a_0}{5120z_{1,0}}-\frac{(2509z_0+1075z_1-3584z_\infty)a_1}{1024z_{1,0}} \nn \\ & \qquad +\frac{(3053z_0-1261z_1-1792z_\infty)a_\infty}{10240z_{1,0}}
\Bigg)
\end{align}
We again note that all dependence on $t_{i,j}$ have canceled, leaving behind a correction term which only depends on the base space information.  We note also that the above excitation $\ell_{-2/5}\ell_{-5/5}$ results in an operator which is not a primary, and this leads to the additional functional dependence on $z_0, z_1, z_\infty$ seen above.  That all of the dependence on the covering space coordinates has completely canceled is no small feat, given the explicit forms of the expansions (\ref{expansionsform2m5}).

Other results that we have obtained are the following:
\begin{align}
&\frac{\langle \left[\ell_{-2/5}\ell_{-2/5}\mathcal{O}_{\infty,-a_\infty/5} \sigma_{(5)}\right](z_\infty)\left[\mathcal{O}_{1,-a_1/2} \sigma_{(2)}\right](z_1)\left[\mathcal{O}_{0,-a_0/4} \sigma_{(4)}\right](z_0)\rangle}{\langle \left[\mathcal{O}_{\infty,-a_\infty/5} \sigma_{(5)}\right](z_\infty)\left[\mathcal{O}_{1,-a_1/2} \sigma_{(2)}\right](z_1)\left[\mathcal{O}_{0,-a_0/4} \sigma_{(4)}\right](z_0)\rangle}\nn \\
&=  \frac{\left(-1/4\right)^{-4/5}\left(\frac{z_{\infty,1}z_{\infty,0}}{z_{1,0}}\right)^{-4/5}}{5^2} \nn \\
& \times \Bigg(\frac{c^2}{256}-\frac{19}{256}c+\frac{a_{\infty}c}{128}-\frac{5a_1 c}{32}+ \frac{a_0 c}{32}+\frac{a_\infty^2}{256}+\frac{25 a_1^2}{16}+\frac{a_0^2}{16}-\frac{5a_0a_1}{8}+\frac{a_0a_\infty}{32}-\frac{5 a_\infty a_1}{32} \nn \\
&\qquad \qquad +\frac{11a_0}{16}-\frac{13a_1}{16}-\frac{125a_\infty}{128}
\Bigg)
\end{align}
\begin{align}
&\frac{\langle \left[\ell_{-7/5}\mathcal{O}_{\infty,-a_\infty/5} \sigma_{(5)}\right](z_\infty)\left[\mathcal{O}_{1,-a_1/2} \sigma_{(2)}\right](z_1)\left[\mathcal{O}_{0,-a_0/4} \sigma_{(4)}\right](z_0)\rangle}{\langle \left[\mathcal{O}_{\infty,-a_\infty/5} \sigma_{(5)}\right](z_\infty)\left[\mathcal{O}_{1,-a_1/2} \sigma_{(2)}\right](z_1)\left[\mathcal{O}_{0,-a_0/4} \sigma_{(4)}\right](z_0)\rangle}\nn \\
&=  \frac{\left(-1/4\right)^{-7/5}\left(\frac{z_{\infty,1}z_{\infty,0}}{z_{1,0}}\right)^{-7/5}}{5} \nn \\
& \times \frac{1}{z_{1,0}}\Bigg(
-\frac{(233z_0-1001z_1+768z_\infty)a_0}{5120}-\frac{(623z_0+148z_1-768z_\infty)a_1}{1024} \nn \\
& \qquad +\frac{(911z_0-527z_1-348z_\infty)a_\infty}{10240}
-\frac{3(53z_0-181z_1+128z_\infty)c}{10240}
\Bigg)
\end{align}
One can use the fact that $\ell_{-5/5}$ acts as a derivative on the simple corrlators we have written above, and so
\begin{align}
&\langle \left[\ell_{-5/5}\ell_{-2/5}\mathcal{O}_{\infty,-a_\infty/5} \sigma_{(5)}\right](z_\infty)\left[\mathcal{O}_{1,-a_1/2} \sigma_{(2)}\right](z_1)\left[\mathcal{O}_{0,-a_0/4} \sigma_{(4)}\right](z_0)\rangle \nn \\
&=\bigg(\frac{-(17/160)c-(1/5)a_\infty-2/5-(1/2)a_1+(1/4)a_0}{z_{\infty,1}} \nn \\ &\qquad \qquad +\frac{-(47/160)c-(1/5)a_\infty-2/5+(1/2)a_1-(1/4)a_0}{z_{\infty,0}}\bigg) \nn \\
& \times \langle \left[\ell_{-2/5}\mathcal{O}_{\infty,-a_\infty/5} \sigma_{(5)}\right](z_\infty)\left[\mathcal{O}_{1,-a_1/2} \sigma_{(2)}\right](z_1)\left[\mathcal{O}_{0,-a_0/4} \sigma_{(4)}\right](z_0)\rangle.
\end{align}
With this and previous results, it is not too hard to show that the above results agree with the fractional algebra $[\ell_{-2/5},\ell_{-5/5}]=3/5\ell_{-7/5}$.

\section{Discussion}

In this work we have considered the correlators of twist operators in symmetric group $S_N$ orbifold CFTs at large $N$, and considered how to compute correlators of descendants using Virasoro generators $L_{m}$ and fractional Virasoro generators $\ell_{m/n_i}$. We have shown that the appropriate ancestors to consider are those that lift to primary operators on the cover, necessarily making them the lifts of a specific class of primary operators in the orbifold theory.  Starting with this special class of primaries, we have shown how covering space techniques may be used to calculate the correlators of descendants given the correlators of the ancestors.  The outline of the techniques is quite algorighmic:
\begin{enumerate}
\item Start with a given $n$-point function with given twist structure, and given operators which lift to primaries on the cover (we have shown that such operators exist, and are necessarily ancestors using weight lowering ladder operators).  This defines the starting $n$-point function.
\item Write a basis of descendants with the operators $L_{m}$ ordered to the left of all $\ell_{m/n_i}$ (all $m<0$).  This is possible because the mixed commutators $[L,\ell]$ close on the $\ell$ operators.  We then set about calculating the descent relationship for each of these basis elements.
\item For a given member of this basis, one eliminates the $L_m$ operators by contour pulls on the base space in the usual way, replacing them with $L_{-1}$ and $L_0$ operators (i.e. derivatives and eigenvalues) acting on lower weight-added correlators and sums of lower weight-added correlators. This process inductively leads to correlators that are a sum of terms with only $\ell$ operators acting at each cycle.  In the process of pulling the contour from one operator, it leads to a sum of $L_m$ modes with $m\geq -1$ acting at other points, for which the $L_0$ and $L_{-1}$ operators may be replaced with the eigenvalues and derivatives.  However, for $m>0$ one simply uses the commutation relations to eliminate these because they annihilate operators that lift to primaries, and the $[L,\ell]$ commutators close on the $\ell$ operators, leaving behind a sum of correlators with the $L$ and $\ell$ operators still correctly ordered with $L$'s to the left.
\item Upon iteration, this leaves correlators with only $\ell$ operators acting, which one may lift to the covering surface (the map to the covering surface is needed for these steps).  Once lifted to the covering space, the $\ell$ operators lift to a sum of Virasoro operators $\mathcal{L}$ on the cover which only appear at ramified points in the map.  However, this sum is necessarily finite, as it acts on some lowest weight module.  Thus, the resulting computation on the cover is necessarily a finite sum of Virasoro operators acting on the appropriate $n'$-point function on the cover.  Thus, one may perform contour pulls on the cover, resulting in an answer which depends on covering space correlator of primaries and covering space derivatives.  For a given correlator, these derivatives are calculable.  The initial correlator is defined using operators which make sense on the base space, so we expect the covering space data to cancel out and not appear in final answers.
\end{enumerate}
To check the last statement above, we perform a proof of principle by calculating some 3-point functions with example descendants.  In our example calculations we see that our expectations are met, and that the descent relations are rather algebraic, and do not depend on the specifics of the seed CFT used, nor the exact operators involved in the correlator.  This feature is particularly important because any sums over orbifold images to make orbifold invariant correlators will each have the same descent relation coefficient.  Thus, the descent relation coefficient is necessarily the same for the orbifold invariant summed form of the operators as well, at least in the simple cases we have seen here.

The algebraic nature of the results is an interesting point, worth deeper investigation.  In the main text we noted that there are obstructions to pulling contours associated with fractional modes.  In the base space this is seen by the fields being multi-valued {\it and} branch cuts exist in the measure functions of the contour integrals.  Going to the covering surface alleviates one of the problems, as the fields on the covering surface are single valued.  However, the branch cuts still exist in the measure, which obstruct simple contour pulls on the cover.  It is only after expanding near the adapted point where an $n^{\rm th}$ root branch cut is cancelled by the function behaving like $(t-t_i)^{n_i}$, and then truncating this series to the appropriate level which makes the contour pulls possible.  The problem of pulling these contours in the covering surface has an interpretation in the base space, as the image of any given contour on the cover can simply be mapped back to the base space, sometimes with several overlapping images, as they are when defining the fractional modes near a twist.  When deforming the contour on the cover, these images no longer must coincide in the base space, and so the single contour on the base dissociates into separate contours, one for each of the copies used to make the twisted field \footnote{Another way of saying this is that the fractional modes are defined by contours in the base space that start with single copy, and wind multiple times around the twist, picking up phases from the measure of the contour integral, and changes of copy index from crossing the cuts that define the simply connected patch \cite{Lunin:2000yv}}.  It would be interesting if there were also a base space interpretation of the truncation of the series on the cover as well.  This could allow for a more direct analysis in the base space, and give a better explanation of how the covering surface information disappears in the final results, possibly through bypassing the covering surface altogether.  A direct way of investigating this would be to use the weight reconstruction algorithm.  Each term in the sum on the cover that was removed corresponds to a covering space operator 
${\mc L}_{m}$ with an $m$ that is too large (it annihilates the operator in question).  These could be reconstructed in terms of the fractional $\ell$ operators on the base, and so upon summing could reconstruct the part of the operator that is zero, and leads to the contour pulls being possible on the cover, giving strong hints as to how the dissociated contours on the base may be able to be pulled separately as well.  Furthermore, the truncation level is known given the gross features of the correlator in question (i.e. the size of the cycle, conformal weight of the operator and list of excitations to the right of the $\ell$ operator which is being truncated once lifted to the cover), making this at least plausible that such an investigation would lead to such algebraic results.  A good place to start would be the covering space global Ward identities, which provide a set of objects that are identically zero on the cover, leading to a set of operators one can construct on the base that one can subtract from correlation functions.

There are also more immediate follow ups to our work presented here, which we are currently exploring.  There are certain cases where the covering space map is already known.  For example, the general $(m)-(n)-(q)$ single cycle covering space map is known in terms of Jacobi polynomials \cite{Lunin:2000yv}.  While these functions are somewhat complicated, they may be treated exactly using recurrence and differential relations amongst Jacobi polynomials.  One may also wish to see how these techniques apply in the cases of higher point functions (our examples focused only on the 3-point functions for brevity).  While the three-point functions are fundamental, providing the structure constants of the theory, the four-point functions allow one to check that all non-trivial structure constants have been found.  There are many such examples of covering space maps for four-point functions \cite{Lunin:2000yv,Pakman:2009zz,Roumpedakis:2018tdb,Dei:2019iym,AlvesLima:2022elo}.  Such investigations are critical when considering the shift in spectrum of operators in the presence of a deformation, as all fusions in the presence of a deformation operator between primary operators at the same weight must be considered: the exchange channels of a 4-point function gives access to all of these at once.  Of course our ultimate goal is to consider situations which have gravitational duals, such as the D1-D5 orbifold CFT.  We see little obstruction to applying the above techniques for various superconformal algebras, where the relevant ancestor operators are presumably those that lift to superconformal primary operators on the cover.  Finally, descent relations such as those mentioned here may aid the process of deforming orbifold CFTs when the deformation ``blow up'' operator is in the twist sector, as it is for the D1-D5 CFT.

Other avenues of investigation also exist.  For example, one may consider more general permutation orbifolds \cite{Haehl:2014yla,Belin:2014fna}, where the group action of the orbifold is realized by a subset of the permutation group acting on copies of a seed CFT.  In this case we see no obstruction, as the permutation orbifold structure admits covering space treatment: patches on the cover directly correspond to copies of the CFT.  Our results here may be thought to agree with the non-chaotic behavior found for such orbifold theories \cite{Belin:2017jli}

One may also ask whether it is possible to relax the large $N$ limit.  In such a case, other genus covering surfaces become important \cite{Lunin:2000yv}, and one still expects to be able to find covering space maps according to the Riemann-Hurwitz formula
\begin{equation}
g=\frac{1}{2}\sum_{i} (n_i-1) -s+1
\end{equation}
where $g$ is the genus of the cover, $n_i$ are the list of cycles in the correlator, and $s$ is the total number of sheets in the cover.  However, the partition function on higher genus surfaces depends on more details of the spectral information of the seed theory, making the covering space techniques of \cite{Lunin:2000yv} for the bare twists more complicated (see for example \cite{Pakman:2009ab} where extremal correlators were considered).  Nevertheless, once these Liouville-dressing-factors are calculated, contour pulls may again be performed on the covering surface.  These do not seem to be sensitive to {\it which} compact surface the correlator is lifted to, although it is difficult to say because this information is indeed encoded in the covering space maps.  However, the ultimate cancellation of the covering space information is still expected such that the answers make sense in the base space CFT.  Seeing how this happens would make these computations interesting as well.

\section*{Acknowledgements}

BAB and AWP wish to thank Ida G. Zadeh for feedback and suggestions on earlier versions of this work.  BAB is thankful for funding support from Hofstra Univeristy including startup funds and faculty research and development grants, and for support from the Scholars program at KITP, which is supported in part by the National Science Foundation under Grant No. NSF PHY-1748958.  The work of AWP is supported by a Discovery Grant from the Natural Sciences and Engineering Research Council of Canada.



\end{document}